\documentclass[10pt,aps,prd,tightenlines,letterpaper,superscriptaddress,nofootinbib,twocolumn]{revtex4-2}

\usepackage{amsmath,latexsym,amsbsy,amssymb,bm,bbold,hyperref}
\usepackage{graphicx}
\usepackage{rsfso}

\usepackage{xcolor}

\usepackage{cancel}

\begin{document}

\def\a{{\alpha}}
\def\be{{\beta}}
\def\d{{\delta}}
\def\D{{\Delta}}
\def\P{{\Pi}}
\def\p{{\pi}}
\def\e{{\varepsilon}}
\def\ep{{\epsilon}}
\def\G{{\Gamma}}
\def\g{{\gamma}}
\def\k{{\kappa}}
\def\l{{\lambda}}
\def\L{{\Lambda}}
\def\m{{\mu}}
\def\n{{\nu}}
\def\o{{\omega}}
\def\O{{\Omega}}
\def\r{{\rho}}
\def\S{{\Sigma}}
\def\s{{\sigma}}
\def\t{{\tau}}
\def\x{{\xi}}
\def\X{{\Xi}}
\def\z{{\zeta}}

\def\ol#1{{\overline{#1}}}
\def\c#1{{\mathcal{#1}}}
\def\b#1{{\bm{#1}}}
\def\eqref#1{{(\ref{#1})}}
\def\wt#1{{\widetilde{#1}}}

\def\ed#1{{\textcolor{magenta}{#1}}}
\def\edd#1{{\textcolor{cyan}{#1}}}

\title{Chiral Symmetry Breaking and Pion Decay in a Magnetic Field}

\author{Prabal~Adhikari}
\email{prabal.adhikari.physics@proton.me}
\affiliation{
        Physics Department, 
        Faculty of Natural Sciences and Mathematics,
        St.~Olaf College,
        Northfield, 
        MN 55057, USA
        }
\affiliation{
	Kavli Institute for Theoretical Physics, 
	University of California,
	Santa Barbara, 
	CA 93106, 
	USA
	}

\author{Brian~C.~Tiburzi}
\email{btiburzi@ccny.cuny.edu}
\affiliation{
	Department of Physics,
        The City College of New York,
        New York,
        NY 10031, USA
        }
\affiliation{
	Graduate School and University Center,
        The City University of New York,
        New York, 
        NY 10016, 
        USA}

\begin{abstract}
The pattern of chiral symmetry breaking is exploited to compute vector and axial-vector pion matrix elements 
in a uniform magnetic field. 
Our results are model independent, and thereby constitute low-energy theorems that must be obeyed by QCD
in external magnetic fields. 
Chiral perturbation theory, lattice QCD and Nambu-Jona-Lasinio results are compared. 
While there is some tension between chiral perturbation theory and lattice QCD, the tension between low-energy QCD and the Nambu-Jona-Lasinio model is more acute.
As an application, 
the matrix elements are utilized to compute pion decay rates in a magnetic field. 
\end{abstract}

\maketitle

\section{Introduction}
\label{introduction}

The impact of magnetic fields on strongly interacting matter is important both theoretically and phenomenologically%
~\cite{Kharzeev:2012ph,Andersen:2014xxa,Miransky:2015ava}. 
Magnetic fields alter the phase diagram of QCD. 
A prime example is magnetic catalysis, 
which enhances chiral symmetry breaking as characterized by the chiral condensate%
~\cite{Shushpanov:1997sf,Agasian:1999sx,Cohen:2007bt}. 
Additionally, 
they alter the nature of the confinement-deconfinement transition, 
which at weak magnetic fields is a crossover. 
For large magnetic fields, 
QCD becomes an anisotropic pure gauge theory that is known to possess a first-order deconfinement transition%
~\cite{Cohen:2013zja}. 
Magnetic fields also modify the phase diagram of QCD at finite isospin and baryon densities through the formation of magnetic vortices and chiral soliton lattices%
~\cite{Adhikari:2015wva,Adhikari:2018fwm,Adhikari:2022cks,Brauner:2016pko,Brauner:2021sci,Evans:2022hwr}. 
There is a wide array of studies exploring the magnetic field axis of the QCD phase diagram
and various order parameters%
~\cite{DElia:2010abb,DElia:2011koc,Bali:2011qj,Bali:2012zg,Bali:2012jv,Bali:2013esa,
Bruckmann:2013oba,Bonati:2013lca,Bonati:2013vba,Bali:2013owa,Bornyakov:2013eya,
Bali:2014kia,Endrodi:2015oba,DElia:2018xwo,Bali:2020bcn,
Ding:2020inp,Ding:2020hxw,DElia:2021yvk,Ding:2022tqn}.

Phenomenologically, 
magnetic fields play an important role in astrophysical scenarios and at high-energy colliders. 
Magnetars support large surface magnetic fields
$\sim 10^{10}$ T, 
and even larger fields are expected in the interior of these neutron stars%
~\cite{Harding:2006qn}.  
Magnetic fields of comparable sizes are possible in the cosmological electroweak phase transition from a 
quark-gluon soup to the confined hadronic phase following the Big Bang%
~\cite{Vachaspati:1991nm,Vachaspati:2020blt}. 
In addition, 
off-central collisions of heavy ions at RHIC and the LHC produce considerably larger magnetic fields of size 
$\sim 10^{16}$ T%
~\cite{Kharzeev:2007jp,Skokov:2009qp,Deng:2012pc,Inghirami:2019mkc,PhysRevX.14.011028}.

Chiral perturbation theory%
~\cite{Gasser:1983yg,Gasser:1984gg} 
is an effective theory that allows for the study of low-energy QCD in a model-independent way, 
including the impact of magnetic fields of size 
$e B \sim m_\p^2$, 
where 
$m_\p$
is the pion mass. 
As referenced above, 
it has been used to predict magnetic catalysis and the phase diagram of finite density QCD. 
Further work has addressed the impact of magnetic fields 
on pions and other hadrons, 
thermodynamic quantities, 
and finite-volume effects relevant for lattice QCD calculations%
~\cite{Werbos:2007ym,Tiburzi:2008ma,Andersen:2012dz,Andersen:2012zc,
Tiburzi:2014zva,Deshmukh:2017ciw,
Hofmann:2020dvz,Hofmann:2020ism,Hofmann:2020lfp,
Adhikari:2021lbl,Adhikari:2021jff,Adhikari:2023fdl}.
In this work, 
we focus on the systematic computation of the left-handed vector current in chiral perturbation theory. 
These contributions are interesting from the perspective of low-energy theorems in a magnetic field, 
while the main application of the result is to leptonic weak decays of mesons.

In a strong magnetic field, 
the charged pion becomes less stable due to an increase in its energy from Landau quantization.%
\footnote{
Due to its magnetic moment, 
the ground-state energy of the outgoing lepton is almost independent of the magnetic field
due to a near zero mode. 
}
Understanding pion stability may be important for magnetar physics, 
for example, 
through its influence on the high-energy neutrino spectrum produced from photo-meson processes%
~\cite{Zhang:2002xv}. 
Weak decay of mesons was the subject of a pioneering lattice QCD calculation%
~\cite{Bali:2018sey}.
A complete model-independent parametrization of the amplitudes entering the 
weak-decay matrix elements in a magnetic field was given in 
Refs.~\cite{Coppola:2018ygv,Coppola:2019wvh}. 
Additional work on the subject has employed models, 
for which an incomplete list would be%
~\cite{Fayazbakhsh:2012vr,Fayazbakhsh:2013cha,Kamikado:2013pya,Simonov:2015xta,Avancini:2016fgq,GomezDumm:2017jij,Andreichikov:2018wrc,Coppola:2019uyr,Coppola:2019idh}.  
Here, 
our focus is exclusively on results that are consequences of low-energy QCD. 
There is an important distinction to be made between model insensitivity and model independency. 
While chiral perturbation theory restricts us to 
$e B \sim m_\p^2$, 
the behavior in this regime provides stringent tests of both hadronic models and lattice QCD computations.

Our presentation has the following organization.
Chiral perturbation theory in a magnetic field is reviewed in 
Sec.~\ref{s:Ch+B}. 
The left-handed vector current is obtained at next-to-leading order in the chiral expansion. 
As detailed in Appendix~\ref{s:L6}, 
an additional tree-level contribution is obtained at next-to-next-to-leading order.  
The resulting vector and axial-vector currents are discussed along with low-energy theorems.  
A few comparisons are made with the pioneering lattice QCD calculation of 
Ref.~\cite{Bali:2018sey}, 
and with the Nambu--Jona-Lasinio model study carried out in 
Ref.~\cite{Coppola:2019uyr}.
As an application, 
the pion current matrix elements are then employed in 
Sec.~\ref{s:decays}
to calculate pion decay rates in a uniform magnetic field. 
For the neutral pion, 
we investigate its electromagnetic decay. 
While a new anomaly-mediated decay mechanism exists, 
the two-photon decay mode remains dominant. 
For the charged pion, 
its weak leptonic decay rate is explicitly shown to respect the residual
$SO(1,1) \times SO(2)$
Lorentz covariance. 
The neutrino angular asymmetry is used to explore the differential decay rate
of longitudinally boosted pions.
A final section, 
Sec.~\ref{s:sum}, 
provides a summary of the main results, 
and concludes with a few avenues for further investigation.

\section{Chiral Lagrangian in a Magnetic Field}
\label{s:Ch+B}

To compute the weak decay rate of pions, 
we obtain the effective left-handed vector current of chiral perturbation theory. 
We use the two-flavor chiral Lagrangian%
~\cite{Gasser:1983yg}
in the presence of an external magnetic field, 
which shares similarities with chiral perturbation theory with the inclusion of virtual photons%
~\cite{Urech:1994hd}.
Properties of the derived vector and axial-vector currents, 
as well as low-energy theorems, 
are then detailed.

\subsection{Leading Order}

At leading order in the chiral expansion, 
the Lagrangian density is 
\begin{equation}
\c L_2
=
\frac{F^2}{4} 
\big\langle \c D^\m U^\dagger \c D_\m U + U^\dagger \chi {+} \chi^\dagger U 
\big\rangle
\label{eq:L2}
,\end{equation}
where angled brackets denote isospin traces 
and 
pions are embedded in the coset field 
$U$
through
$U = \exp ( i \bm{\phi} \cdot \bm{\tau} / F )$, 
with 
$\tau^a$
as the isospin matrices. 
The canonically normalized pion fields are
$\p^\pm = \frac{1}{\sqrt{2}} \left( \phi^1 \mp i \phi^2 \right)$ 
and
$\p^0 = \phi^3$, 
which we will often abbreviate as
$\p^Q$, 
for 
$Q = \pm 1$
and 
$0$.  
The parameter
$F$
is the chiral-limit value of the pion-decay constant in vanishing magnetic field. 
External scalar 
$s$
and pseudoscalar 
$p$
sources are contained in the spurion field
$\chi = 2 B_0 (s + i p)$, 
where the parameter 
$B_0$
is related to the chiral-limit value of the chiral condensate
$\langle \ol \psi \psi \rangle_0 = - 2 B_0 F^2$.  
With the scalar source set equal to the quark mass 
$m_q$
for vanishing pseudoscalar source, 
we obtain the 
Gell-Mann--Oakes--Renner relation
$m^2 F^2 = - \langle \ol \psi \psi \rangle_0  \, m_q$, 
where 
$m$
is the tree-level pion mass. 
The vector source fields appear in 
Eq.~\eqref{eq:L2}
through the action of the gauge covariant derivative 
\begin{equation}
\c D^\m U = \partial^\m U + i U L^\m - i R^\m U
,\end{equation}
where 
$L^\m$
is a left-handed vector field, 
and 
$R^\m$
is a right-handed vector field.

The left-handed current is obtained from differentiation with respect to the left-handed source field, 
which we decompose into isoscalar and isovector terms 
$L^\m =  \frac{1}{2} \langle L^{\m} \rangle  \, \mathbb{1} + \sum\limits_{a = 1}^3 L^{\m a}  \tau^a $. 
The isovector left-handed current in a background electromagnetic field is then
\begin{equation}
J_L^{\m a}
=
\frac{\partial \c L}{\partial L^a_\m} \, \Bigg|_{L_\m = R_\m = - e Q A_\m}
\label{eq:JL}
,\end{equation}
where 
$e >0$
is the magnitude of the electron's charge, 
and 
\begin{equation}
Q = \frac{1}{6} \, \mathbb{1} + \frac{1}{2} \tau^3
,\end{equation}
is the light quark charge matrix.  
The vector potential of the electromagnetic field is denoted by 
$A_\m$, 
and we implement the magnetic field 
$\b B  = B \, \b{\hat{z}}$
using the asymmetric gauge 
$A_\m(x) = (0, - B y, 0, 0)$, 
and choose 
$B > 0$. 
With the magnetic field included in 
Eq.~\eqref{eq:L2}, 
the power counting is
\begin{equation}
p^2 \sim m_\p^2 \sim e B \ll \L_\chi
,\end{equation} 
where 
$p$
is a good component of momentum and
$\L_\chi = 4 \p F_\p = 1.16 \, \texttt{GeV}$
is the chiral symmetry breaking scale.  
While the magnetic field must be perturbatively small in the chiral expansion, 
this power counting accommodates fields that are large in physical terms, 
because 
$e B / m_\p^2 \sim 1$.

The leading-order (LO) isovector left-handed current obtained from 
Eq.~\eqref{eq:L2}
is simply
\begin{equation}
\left( J_L^{\m Q} \right)_{\text{LO}}
=
F \, D^\m \p^Q
\label{eq:JLO}
,\end{equation}
where 
$J_L^{\m Q}$
is shorthand for 
$\frac{1}{\sqrt{2}} \left( J_L^{\m 1} \pm i J_L^{\m 2} \right)$
for 
$Q = \pm 1$, 
and
$J_L^{\m 3}$
for 
$Q = 0$. 
The electromagnetic gauge covariant derivative of the pion field is specified by 
\begin{equation}
D^\m \p^Q = \left( \partial^\m + i Q e A^\m \right) \p^Q
.\end{equation} 
As 
$\c L_2$
has even intrinsic parity, 
the LO contribution to the left-handed current is entirely axial vector in nature. 
For neutral pions, 
there is no effect from an electromagnetic field at 
LO; 
whereas, 
for charged pions, 
the effect is due to minimal coupling with the electromagnetic field.

\subsection{Next-To-Leading Order}

At next-to-leading order
(NLO) 
in the chiral expansion, 
there are several effects to account for. 
Expanding 
Eq.~\eqref{eq:L2}
to 
NLO 
generates one-loop diagrams that modify the effective action. 
Additionally, 
there are terms required from the 
NLO 
chiral Lagrangian
$\c L_4$. 
These include counter-terms necessary to renormalize the one-loop diagrams, 
but there are also finite contributions. 
The NLO effects are categorized below, 
with results given for both neutral and charged pions.

\subsubsection{Mass Renormalization}

The four-pion terms obtained by expanding 
Eq.~\eqref{eq:L2}
to 
NLO
generate tadpole diagrams that result in 
wavefunction and mass renormalization. 
There is also the mass counter-term
$l_3$
from the 
NLO 
chiral Lagrangian, 
which cancels the ultraviolet divergence contained in the one-loop pion mass.  
As this counter-term is independent of the magnetic field, 
the renormalized magnetic mass is identical to that obtained from a simple zero-field subtraction. 
For the neutral pion, 
one obtains%
~\cite{Tiburzi:2008ma}
\begin{equation}
m_{\p^0}^2(B)
= 
m_{\p}^2
\left[
1 + \frac{eB}{\L_\chi^2} \,
\c I \left( \frac{m_\p^2}{eB} \right)
\right]
\label{eq:mpi0}
,\end{equation}
where the function 
$\c I(z)$
arises from the charged-pion tadpole, 
and is given by
\begin{equation}
\c I(z)
=
2 \log \G\left(\frac{1+z}{2}\right)
+
z \left( 1 - \log \frac{z}{2} \right)
- 
\log 2 \p
\label{eq:I(z)}
,\end{equation}
with 
$\c I(z) = - \frac{1}{6} z^{-1} + \c O(z^{-3})$
for 
$z \gg 1$.
The physical pion mass 
$m_\p$
and physical decay constant
$F_\p$
appear in Eq.~\eqref{eq:mpi0}
up to corrections that are of 
next-to-next-to-leading order
(NNLO).

For the charged pion, 
the one-loop contribution to its mass 
arises only from a neutral pion tadpole, 
after wavefunction renormalization is accounted for. 
This contribution is thus independent of the magnetic field to 
NNLO. 
There are, 
however, 
counter-terms that depend on external fields contained in 
the 
NLO 
chiral Lagrangian%
~\cite{Gasser:1983yg}
\begin{multline}
\c L_4 
\supset
l_5 
\langle U^\dagger \hat{R}^{\m \n} U \hat{L}_{\m \n} \rangle
\\+ 
l_6 \tfrac{i}{2}
\langle \c D^\m U^\dagger \c D^\n U \hat{L}_{\m \n} + \c D^\m U \c D^\n U^\dagger \hat{R}_{\m \n} \rangle
\label{eq:L4}
,\end{multline} 
where the left- and right-handed field-strength tensor are defined by
\begin{eqnarray}
L^{\m \n} 
&=& 
\partial^\m L^\n - \partial^\n L^\m - i [L^\m, L^\n]
,\notag \\
R^{\m \n} 
&=& 
\partial^\m R^\n - \partial^\n R^\m - i [R^\m, R^\n]
\label{eq:LnR}
.\end{eqnarray}
The hats are employed to denote trace-subtracted quantities, 
so that for any matrix 
$O$, 
we have
\begin{equation}
\hat{O} = O - \frac{1}{2} \langle O \rangle \, \mathbb{1}
.\end{equation} 
The terms from 
$\c L_4$
contribute to the magnetic mass-squared of the charged pion%
\footnote{
As is made clear in 
Eq.~\eqref{eq:Epion}
below, 
the magnetic mass of the charged pion is defined to be the contribution to its energy 
that is independent of the Landau level quantum number.
} 
\begin{equation}
m_{\p^\pm}^2(B)
=
m_\p^2
\left[
1
+ 
\overline{l\phantom{l}}
\, 
\frac{eB}{m_\p^2}
\,
\frac{eB}{\L_\chi^2}
\right]
\label{eq:mpipm}
,\end{equation}
and encompasses its magnetic polarizability. 
We use the abbreviation 
\begin{equation}
\overline{l\phantom{l}}  
\equiv 
\tfrac{1}{3} \left(
\overline{l_6}
 - 
\overline{l_5} \,
\right)
=
1.0 \pm 0.1
\label{eq:LEC=1}
,\end{equation}
for the linear combination of renormalized low-energy constants. 
This combination is renormalization scale and scheme independent
(the chiral logarithm cancels in the difference), 
and the value quoted is obtained from an NNLO analysis of rare pion decays%
~\cite{Bijnens:2014lea}.
\subsubsection{Magnetic Polarizability \& Comparison with Nambu-Jona-Lasinio Model}
\label{sec:NJL}
For weak magnetic fields, the neutral pion mass in a magnetic field is smaller than the zero-field mass, $m_{\p^{0}}(B)<m_{\p^{0}}$ as is evident upon expanding in powers of the magnetic field
\begin{equation}
m_{\pi^0}(B)
= 
m_\pi - \frac{1}{2} \beta^{\pi^0}_M B^2 + \mathcal{O}(B^4)\ .
\end{equation}
The neutral pion polarizability, $\beta^{\pi^0}_{M}$,
\begin{equation}
\beta^{\pi^0}_M
=
\frac{e^2}{6 m_\pi (4 \pi F_\pi)^2}
,\end{equation}
admits a drastically different structure from that in the Nambu-Jona-Lasinio analysis, 
\begin{equation}
\frac{\left.\beta_M^{\pi^0}\right|_{\textrm{NJL}}}{\beta_M^{\pi^0}}
=
\left( \frac{F_\pi}{M} \right)^2 \frac{20/9}{- I_2(M,\Lambda)}
+ 
\mathcal{O} \left( \frac{m_\pi^2}{4 M^2} \right)
\approx 
\frac{1}{3}
.\end{equation} 
Here, there is dependence on the constituent quark mass, $M$. and the ultraviolet cutoff, $\Lambda$, but in a renormalization group non-invariant manner. 
Since these are model-dependent parameters, we utilize Set I values and the loop function $I_{2}$ of Ref.~\cite{Coppola:2018vkw}, resulting in an estimate for the neutral pion polarizability that is qualitatively in agreement but approximately a factor of three smaller.

A similar analysis for the charged pion leads to a polarizability value of
\begin{align}
\beta_M^{\pi^{\pm}}
=
- \frac{e^2 \, \overline{\ell}}{m_\pi (4 \pi F_\pi)^2}\,.
\end{align}
Comparison with the Nambu-Jona-Lasinio model produces the result that is not only quantitatively in disagreement,
\begin{align}
\frac{\left.\beta_M^{\pi^\pm}\right|_{\textrm{NJL}}}{\beta_M^{\pi^\pm}}
&=
\left( \frac{F_\pi}{M} \right)^2 \frac{17/27}{- I_2(M,\Lambda)}
+ 
\mathcal{O} \left( \frac{m_\pi^2}{4 M^2} \right)\\
&\approx 
- \frac{1}{10},
\end{align}
but also produces the wrong qualitative behavior compared with chiral perturbation theory.
\subsubsection{Left-Handed Current Renormalization}

In expanding 
Eq.~\eqref{eq:L2}
to NLO, 
there are three-pion terms%
\footnote{
There are also two-pion terms with one left-handed source, 
but these do not contribute to pion decay. 
The origin of such terms is explained by writting
$L^\m = V^\m - \c A^\m$, 
where 
$V^\m$
is a vector source and 
$\c A^\m$
is an axial-vector source. 
From 
$\c L_2$, 
which has even intrinsic parity, 
the terms with 
$V^\m$
have an even number of pions, 
while those with  
$\c A^\m$
have an odd number of pions. 
}
accompanied by one left-handed vector source
$L^\m$. 
These terms generate one-loop corrections to the left-handed current.  
Taken with the wavefunction renormalization, 
the divergence in these tadpole diagrams is canceled by the 
$l_4$
counter-term from 
$\c L_4$, 
which is independent of the magnetic field.
Accordingly, 
the renormalized one-loop result in a magnetic field is identical to that obtained from 
a zero-field subtraction. 
The result is a renormalization of the leading-order current
Eq.~\eqref{eq:JLO}, 
where 
$F$
is replaced by%
~\cite{Tiburzi:2008ma} 
\begin{equation}
F_{\p^Q}(B)
=
F_\p
\left[ 1 - \left( 1 - \frac{|Q|}{2} \right) \frac{e B}{\L_\chi^2} \c I \left( \frac{m_\p^2}{eB} \right) \right]
\label{eq:FpiB}
,\end{equation}
where the tadpole function appears in 
Eq.~\eqref{eq:I(z)}.

The external-field dependent terms appearing in 
$\c L_4$
that are written in 
Eq.~\eqref{eq:L4}
also contribute to the 
left-handed current at NLO. 
Performing an integration by parts at the level of the action, 
we subsequently obtain from 
Eq.~\eqref{eq:JL}
a NLO contribution 
\begin{equation}
\left(
J_L^{\m Q}
\right)_\text{NLO}
=
- \overline{l\phantom{l}}  
 \, F_\p \frac{i Q e F^{\m \n}}{\L_\chi^2} 
 D_\n \, \p^Q
,\end{equation}
where
$\overline{l\phantom{l}}$
is the linear combination of low-energy constants given in 
Eq.~\eqref{eq:LEC=1}. 
This contribution to the left-handed current has axial-vector quantum numbers; 
and, 
it vanishes for the neutral pion due to charge conjugation.

The final contribution to the left-handed current at NLO
arises from the chiral anomaly%
~\cite{Adler:1969gk,Bell:1969ts}. 
The Wess-Zumino-Witten 
(WZW)
Lagrangian%
~\cite{Wess:1971yu,Witten:1983tw}
has been constructed directly in two-flavor chiral perturbation theory in 
Ref.~\cite{Kaiser:2000ck}. 
Retaining only the terms relevant for the left-handed current, 
we have
\begin{multline}
\c L_4^{\text{WZW}}
\supset
\frac{N_c}{64 \p^2}
\e^{\m \n \a \be}
\Bigg\langle
i \left( 
U^\dagger \partial_\m U \hat{L}_\n
-
U \partial_\m U^\dagger \hat{R}_\n
\right)
\\
+ 
U^\dagger [ \hat{R}_\m, U] \hat{L}_\n
\Bigg\rangle
\, \big\langle L_{\a \be} {+} R_{\a \be} \big\rangle
\label{eq:WZW}
,\end{multline}
with the convention that 
$\e^{0123} = +1$. 
Note that because a pure 
$\text{SU}(2)$
theory has no anomalies, 
this Lagrangian density vanishes without the inclusion of  isoscalar vector sources. 
Such a source is provided by an external electromagnetic field 
through the isoscalar contribution to the charge matrix.

Appealing to 
Eq.~\eqref{eq:JL}, 
we obtain the final 
NLO  
contribution to the left-handed current using the WZW Lagrangian
\begin{equation}
\left( J_L^{\m Q} \right)_\text{WZW}
=
- \frac{e \widetilde{F}^{\m\n}}{8 \p^2 F_\p} D_\n \, \p^Q
\label{eq:JLWZW}
,\end{equation} 
where the dual electromagnetic field-strength tensor is 
$\widetilde{F}^{\m \n} = \frac{1}{2} \e^{\m \n \a \be} F_{\a \be}$.
This contribution to the current has vector quantum numbers
because 
$\c L_4^\text{WZW}$
has odd intrinsic parity. 
Despite the necessary gauge non-invariance of the 
WZW Lagrangian, 
the anomalous contribution to the left-handed current appropriately maintains electromagnetic gauge covariance.

\subsection{Currents and Low-Energy Theorems}

In a uniform magnetic field, 
general parameterizations of the vector and axial-vector current matrix elements of pions were given in 
Ref.~\cite{Coppola:2018ygv}. 
To facilitate comparison with the NLO results obtained in chiral perturbation theory, 
we write analogous parameterizations in terms of the vector current operator 
\begin{equation}
J_V^{\m Q}
=
- e 
\,
F_{\p^Q}^{(V)}
\, 
\wt F {}^{\m \n} D_\n \, \p^Q
\label{eq:JV}
\end{equation}
and the axial-vector current operator
\begin{multline}
J_A^{\m Q}
=
-
\Bigg[
F_{\p^Q}^{(A1)}
D^\m
- 
i Q e 
\,
F^{(A2)}_{\p^Q}
\,
F^{\m \n} D_\n
\\+
e^2
F^{(A3)}_{\p^Q}
\,
F^{\m \n} F_{\n \a} D^\a
\Bigg]
\p^Q 
\label{eq:JA}
.\end{multline}
The amplitudes 
$F^{(V)}$
and
$F^{(Aj)}$,
for 
$j = 1$--$3$, 
are generally required for pion weak decay in a uniform magnetic field, 
with only 
$F^{(A1)}$
remaining in the zero-field limit.  
Note that the charge label appearing on 
$F^{(A2)}_{\p^Q}$
is superfluous. 
As written, 
it is identical for the charged pions, 
and necessarily drops out of the axial-vector current for neutral pions. 
These are consequences of charge conjugation. 

Using the chiral perturbation theory calculation
Eq.~\eqref{eq:JLWZW}, 
we have the vector amplitudes 
\begin{equation}
F^{(V)}_{\p^\pm}
= 
F^{(V)}_{\p^0}
=
\frac{1}{8 \p^2 F_\p} 
\left[ 
1 
+
\c O \left( \frac{eB}{\L_\chi^2} \right)
\right]
,\end{equation}
where these results are exactly fixed by the chiral anomaly
up to NNLO corrections, 
the size of which has been indicated. 
Note that these corrections could equally well be typified as 
$m_\p^2 / \L_\chi^2$,
because we assume
$e B / m_\p^2 \sim 1$. 
The anomaly as the origin of the vector transition amplitude of a pion in a magnetic field was noted in 
Ref.~\cite{Coppola:2019uyr}.
Lattice QCD calculations of the vector-current matrix element of 
the pion in a background magnetic field therefore enable extraction of the chiral anomaly.%
\footnote{
In Ref.~\cite{Bali:2018sey}, 
the existence of the vector transition amplitude was interpreted in terms of   
$\rho$--$\p$ 
mixing in a magnetic field%
~\cite{Bali:2017ian}. 
While the chiral anomaly thus provides the mechanism for such mixing, 
obtaining a prediction for the mixing angle requires knowledge of the model-dependent coupling  
$g_{\g \rho \p}$
that appears in the Lagrangian density
$\c L = \frac{1}{3} e \, g_{\g \rho \p} \wt F {}^{\m \n} \partial_\m \b \rho_\n \cdot \b \phi $.
In the phenomenologically successful vector-dominance model, 
this coupling vanishes in the chiral limit, 
see, 
for example, 
Ref.~\cite{Harada:2003jx}.
}
The lattice QCD results for 
$F_{\p^\pm}^{(V)}$
approach a constant value in the limit of small magnetic fields%
~\cite{Bali:2018sey}%
\footnote{
Using their notation and conventions 
for the matrix element, 
the chiral anomaly leads to the prediction 
$f'_\p (0) = (4 \p^2 f_\p )^{-1}$.
With the relation
$f_\p = \sqrt{2} \, F_\p$, 
moreover, 
we see that 
$f'_\p (0) \big/ f_\p = F^{(V)}_{\p^\pm} (0) \big/ F_\p$. 
}
\begin{equation}
\frac{F_{\p^\pm}^{(V)}(B{=}0)}{F_\p}
=
\begin{cases}
1.1(2) \, \texttt{GeV}^{-2}
&
\text{Wilson quarks}
\\
0.8(2) \, \texttt{GeV}^{-2}
&
\text{Staggered quarks}
\\
1.49(3) \, \texttt{GeV}^{-2}
&
\text{Anomaly, Eq.~\eqref{eq:JLWZW}}
\end{cases}
\label{eq:anomaly}
,\end{equation}
where the uncertainty in the theory prediction from the chiral anomaly is an estimate of pion-mass 
corrections at
NNLO. 
The lattice data with quenched Wilson quarks are obtained with a larger-than-physical pion mass
$m_\p \approx 415 \, \texttt{GeV}$, 
while the staggered data are obtained at the physical quark masses. 
The two lattice QCD results are consistent with each other. 
The prediction from the chiral anomaly is independent of the quark masses and of dynamical quarks, 
for which the agreement between fermion discretizations might be expected. 
Relative pion mass corrections to the Wilson results are expected to be roughly
$m_\p^2 / \L_\chi^2 \sim 13\%$, 
and the quenched Wilson lattice QCD result is thus completely consistent with the anomaly. 
Taken by itself, 
however,
there is 
$\sim 3.5 \, \sigma$ 
tension between the fully dynamical staggered lattice QCD result at the physical pion mass and the value required by the chiral anomaly. 
\begin{figure}[t]
	\centering 
	\includegraphics[width=0.475\textwidth, angle=0]{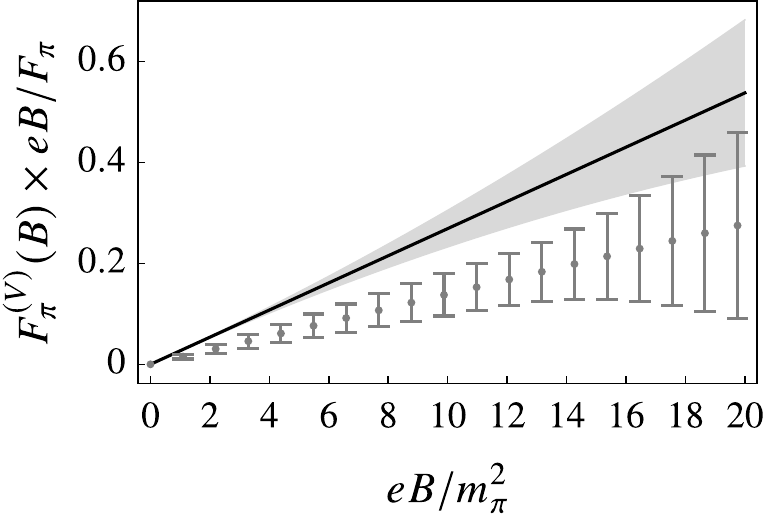}	
	\caption{Comparison of $F^{(V)}$ from chiral perturbation theory with that for staggered dynamic quarks. The bands indicate the size of next-to-leading order corrections.} 
	\label{f:FV}%
\end{figure}
In Fig.~\ref{f:FV}, we plot $F_{\p}^{(V)}$ from chiral perturbation theory and compare with lattice data. Agreement, accounting for uncertainties, is stronger for larger fields. For weaker fields, where predictions of chiral perturbation theory have the lowest uncertainties, there is disagreement.

\begin{figure}[t]
	\centering 
	\includegraphics[width=0.475\textwidth, angle=0]{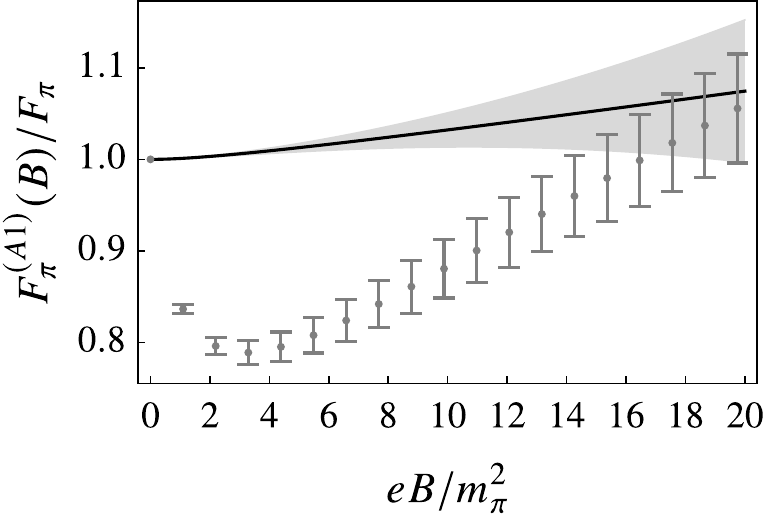}	
	\caption{Comparison of $F^{(A1)}$ from chiral perturbation theory with that for staggered dynamic quarks. The bands indicate the size of next-to-leading order corrections.} 
	\label{f:FA1lattice}%
\end{figure}

\begin{figure}[t]
	\centering 
	\includegraphics[width=0.475\textwidth, angle=0]{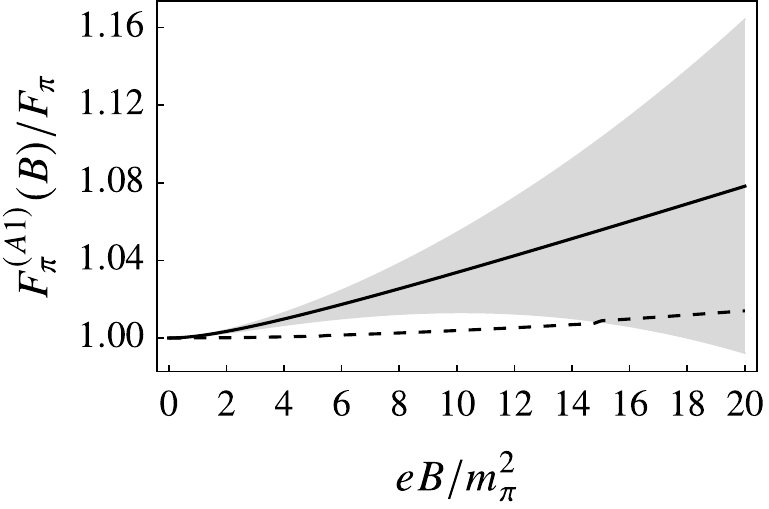}	
	\caption{Comparison of $F^{(A1)}$ from chiral perturbation theory with that from the Nambu-Jona-Lasinio model (dotted line). The bands indicate the size of next-to-leading order corrections.} 
	\label{f:FA1NJL}%
\end{figure}

Returning to Eq.~\eqref{eq:JA}
to address the axial-vector amplitudes, 
the first is simply given by 
$F_{\p^Q}^{(A1)} = F_{\p^Q}(B)$, 
where the latter appears in 
Eq.~\eqref{eq:FpiB}. 
There are lattice QCD results for this amplitude%
~\cite{Bali:2018sey}, 
however, 
the axial-vector data appear to be described by linear 
$e B$
dependence, 
for which a negative slope is found near 
$e B = 0$. 
By contrast, 
the chiral perturbation theory prediction from 
Eq.~\eqref{eq:FpiB}
requires strikingly different behavior
\begin{equation}
F_{\p^\pm}(B)
=
F_\p
\left[
1 
+ 
\frac{ (eB)^2}{12 m_\p^2 \L_\chi^2}
+ 
\cdots 
\right]
\label{eq:FA1small}
,\end{equation}
in the small-field limit 
$e B / m_\p^2 \ll 1$. 
Even for magnetic fields satisfying
$e B \sim m_\p^2$, 
chiral perturbation theory requires 
$F_{\p^\pm}(B) \big/ F_\p > 1$. 
We compare the results in Fig.~\ref{f:FA1lattice}. 
While there is agreement for a limited range of magnetic field $eB\sim 15-20 m_{\p}^{2}$, the disagreement at low magnetic fields is severe. In Fig.~\ref{f:FA1NJL}, we compare chiral perturbation theory with the Nambu-Jona-Lasinio model using Set I values of Ref.~\cite{Coppola:2018vkw} -- the disagreement at weak fields is smaller compared to the disagreement with the lattice.

For the two remaining axial-vector amplitudes in
Eq.~\eqref{eq:JA}, 
chiral perturbation theory requires
\begin{align}
F^{(A2)}_{\p^\pm} 
&=
\frac{\overline{l\phantom{l}}}{16 \p^2 F_\p}
\left[
1 + \c O \left( \frac{eB}{\L_\chi^2} \right)
\right]
\label{eq:FA2}
,\\
F_{\p^Q}^{(A3)} 
&= 
-
\frac{c_{\p^Q}}{F_\p^3}
\left[ 1 + \c O \left( \frac{eB}{\L_\chi^2} \right) \right]
\label{eq:FA3}
.\end{align}
Neither of these axial-vector amplitudes have been calculated with lattice QCD, 
for which the former is predicted from chiral perturbation theory with an uncertainty of 
$10\%$. 
The latter, 
however, 
vanishes at NLO, 
which is easily established from power counting.%
\footnote{
This was pointed out in 
Ref.~\cite{Tiburzi:2008ma}, 
where the decomposition of the axial-vector current 
Eq.~\eqref{eq:JA}
essentially already appears. 
}
The non-vanishing contribution obtained for the
$F_{\p^Q}^{(A3)}$
amplitude thus arises at NNLO, and requires operators contained in the 
$\c L_6$
chiral Lagrangian%
~\cite{Bijnens:1999sh}.
The calculation leading to  
Eq.~\eqref{eq:FA3}
is described in 
Appendix~\ref{s:L6}. 
There are no determinations of the required low-energy constants appearing in the 
$c_{\p^Q}$
parameter, 
although there are resonance saturation estimates for some of the contributing terms
\cite{Cirigliano:2006hb}. 

With the currents 
Eqs.~\eqref{eq:JV} and \eqref{eq:JA}
derived in the effective theory, 
we can investigate conservation laws and additional low-energy theorems. 
The vector current of the neutral pion in an external field is conserved, 
which is easily demonstrated by taking the divergence 
\begin{equation}
\partial_\m J_V^{\m 3}
=
0
,\end{equation}
and using the Maxwell equation 
$\partial_\m \wt F^{\m \n} = 0$. 
For the charged pion, 
the currents are necessarily only gauge covariant, 
for which it is natural to consider the gauge covariant derivative of the currents.  
Using the vector current of the charged pion from 
Eq.~\eqref{eq:JV}, 
we obtain the covariant divergence of the vector current
\begin{equation}
D_\m J_V^{\m \pm}
=
\pm 
2 
i e^2\,
F_{\p^Q}^{(V)} \, \b{E} \cdot \b{B}
\, \p^\pm
.\end{equation}
Hence, 
the vector current of charged pions is covariantly conserved in a uniform magnetic field.

Explicit chiral symmetry breaking is seen from the derivatives of the 
axial-vector currents. 
Taking the divergence of the neutral pion's axial-vector current
Eq.~\eqref{eq:JA}, 
we obtain 
\begin{equation}
\partial_\m J_A^{\m 3}
=
F_{\p^0}(B) \, m_{\p^0}^2(B) \, \p^0
\label{eq:dpi0}
,\end{equation}
for an on-shell $\p^0$. 
Here, 
we have dropped the contribution from the term with
$F_{\p^0}^{(A3)}$, 
as this amplitude is zero at NLO. 
Non-zero quark masses lead to explicit chiral symmetry breaking, 
which manifests in the non-vanishing divergence. 
The divergence of the axial-vector current does vanish in the chiral limit, 
because 
$m_{\p^0}^2(B) \longrightarrow 0$
for fixed magnetic field from 
Eq.~\eqref{eq:mpi0}. 
On the other hand, 
the divergence of the axial-vector--pseudoscalar correlation function contains a pole
\begin{equation}
\int
d^4 x \,
e^{ i x \cdot p}
\big\langle 0 \, |  
\,
\partial_\m J_A^{\m 3} (x) \, \p^0(0) \,
| 0 \, \big\rangle
=  
\frac{i F_{\p^0}(B) \, p^2}{p^2 - m_{\p^0}^2(B) + i \e}
\label{eq:dpi0pole}
.\end{equation}
This divergence does not vanish when one takes the chiral limit before putting the pion on-shell, 
and the corresponding chiral symmetry is spontaneously broken. 
The neutral pion remains a pseudo-Goldstone boson in a magnetic field, 
and a straightforward generalization of the Gell-Mann--Oakes--Renner relation is satisfied. 
This can be confirmed explicitly to NLO by taking the ratio of the relation in a magnetic field to that in zero field
\begin{equation}
\frac{F_{\p^0}^2 (B) \, m_{\p^0}^2(B)}{F_\p^2 \, m_\p^2} 
=
\frac{\langle \ol \psi \psi \rangle_B}{\langle \ol \psi \psi \rangle_0} 
=
1 - \frac{eB}{\L_\chi^2} \c I \left( \frac{m_\p^2}{eB} \right)
\label{eq:cond}
,\end{equation}
where the chiral condensate ratio on the right-hand side was first determined using chiral perturbation theory in
Ref.~\cite{Cohen:2007bt}.

For charged pions, 
the situation is different. 
The gauge covariant derivative of the current leads to
\begin{equation}
D_\m J_A^{\m \pm}
=
-
\left[
F_{\p^{\pm}}^{(A1)}
D_\m D^\m
+ 
\frac{e^2}{2} 
F_{\p^\pm}^{(A2)}
F_{\m \n} F^{\m \n}
\right]
\p^\pm
,\end{equation}
where we have made use of the uniformity of the external field, the commutation relation $[D_{\m},D_{\n}]=ieF_{\m\n}$
and dropped the contribution from 
$F_{\p^\pm}^{(A3)}$
as it vanishes to NLO. Nevertheless, due to the commutation relation, an analogous relation would hold at NNLO.
In the case of charged pions, 
we can still appeal to the Green's function 
\begin{equation}
\Big[ D_\m D^\m + m_{\p^\pm}^2(B) \Big]
G_\pm (x,0)
= 
- i \d^{(4)}(x)
.\end{equation}
Notice that the Green's function is determined at NLO
from the renormalized action, 
in which the charged pion mass has been replaced by the magnetic mass. 
After accounting for the wavefunction renormalization, 
the gauge covariant derivatives are not renormalized.

The un-amputated, 
coordinate-space correlation function between the current's divergence and the charged pion 
can be written in term of the Green's function. 
We find
\begin{multline}
\langle \, 0 \, | D_\m J_A^{\m \pm}(x) \, \p^\pm(0) \, | \, 0 \, \rangle
\\
=
\left[
F_{\p^\pm}^{(A1)}
m^2_{\p^\pm} (B)
-
F_{\p^\pm}^{(A2)}
(e B)^2
\right]
G_\pm(x,0)
,\end{multline}
provided 
$x_\m \neq 0$. 
This expression can be simplified further using the NLO quantities determined in 
Eqs.~\eqref{eq:mpipm}, 
\eqref{eq:FpiB}, 
and 
\eqref{eq:FA2}. 
We find the simple result
\begin{equation}
\langle \, 0 \, | D_\m J_A^{\m \pm}(x) \, \p^\pm(0) \, | \, 0 \, \rangle
=
F_{\p^\pm}(B) \, 
m^2_\p
\,
G_\pm(x,0)
\label{eq:dpipm}
,\end{equation}
up to NNLO corrections.
At the level of the charged pion Green's function,  
the covariant divergence of the axial-vector current vanishes in the chiral limit.
This result, 
however, 
is intermediate to those obtained for the neutral pion in
Eqs.~\eqref{eq:dpi0} and \eqref{eq:dpi0pole}.
The quark mass appears as the source of explicit chiral symmetry breaking, 
however, 
spontaneous symmetry breaking cannot be exhibited for charged pions. 
While the Green's function maintains poles encompassing the Landau levels, 
the corresponding energies do not vanish in the chiral limit. 
Near a pole, 
the correlation function in 
Eq.~\eqref{eq:dpipm}
consequently vanishes in the chiral limit. 
In a magnetic field, 
the charged pion is technically no longer a pseudo-Goldstone boson.%
\footnote{
In the chiral limit, 
the two-flavor QCD action in a magnetic field maintains only the diagonal
$U(1)_L {\times} U(1)_R$
subgroup of the 
$SU(2)_L {\times} SU(2)_R$
chiral symmetry group that exists in vanishing magnetic field. 
The neutral pion emerges as the pseudo-Goldstone boson associated with the spontaneous breaking of 
$U(1)_L {\times} U(1)_R \longrightarrow U(1)_V$.} 
The simplicity of 
Eq.~\eqref{eq:dpipm}
is ultimately a reflection of the underlying QCD relation
\begin{equation} 
D_\m \Big[ \, \ol \psi(x) \g^\m \g^5 \tau^\pm \psi(x) \Big] = (m_u {+} m_d)\, \ol \psi(x) i \g^5 \tau^\pm \psi(x)
\label{eq:QCD}
,\end{equation}
in a magnetic field,  
where 
$D_\m$
is the electromagnetic gauge covariant derivative
and 
$\psi(x) = \begin{pmatrix} u(x) \\ d(x) \end{pmatrix}$
is the isodoublet of light quark fields. 

While the charged pion does not produce a Gell-Mann-Oakes-Renner relation in the spirit of Eq.~\eqref{eq:cond}, it proves worthwhile to consider the coupling of pions to the vacuum through the pseudoscalar density. 
The isovector pseudoscalar current (partially) appearing in Eq.~\eqref{eq:QCD}
can be obtained in the effective theory from differentiation with respect to the corresponding source
\begin{equation}
\label{eq:def-JP}
J_{P}^a = \frac{\partial \c L}{\partial p^a} \, \Bigg|_{\,p=0}
.\end{equation}
Using Eq.~\eqref{eq:L2} in a background magnetic field
(as well as the requisite field-independent counter-term from $\c L_4$), 
we find that the pion pseudoscalar current at NLO can be written as
\begin{align}
J^Q_P
&=
G_{\p^Q}(B)
\, \p^Q \left[ 1 + \c O \left( \frac{eB}{\L_\chi^2} \right) \right]
.\end{align}
The modification of $G_{\p^{Q}}(B)$ in terms of its zero-field counterpart~\cite{Gasser:1983yg}, $G_{\p}$, is
\begin{align}
\label{eq:GB}
G_{\p^{Q}}(B)&=G_{\p}\left[1-\frac{|Q|}{2}\frac{eB}{\L^{2}_{\chi}}\c{I}\left(\frac{m_{\p}^{2}}{eB}\right)\right]
.\end{align}
At zero field, $G_{\p}$, measures the coupling of the pion to the vacuum through the pseudoscalar matrix element,
\begin{align}
\langle\,0\,|\,\ol\psi i \g_{5} \tau^{a}\psi\,|\,\pi^{b}\rangle=G_{\p}\d^{ab}
.\end{align}
It relates to the pion mass and decay constant through 
\begin{align}
F_{\p}m_{\p}^{2}=\tfrac{1}{2}(m_{u}+m_{d})G_{\p}
\end{align}
with the isospin limit assumed -- 
the identity follows from the divergence of the axial current.
A detailed calculation of $G_{\p^Q}$ in a magnetic background is presented in Appendix~\ref{s:JP}.
Upon computing the matrix element of Eq.~\eqref{eq:QCD} with a charged pion field and amputating the Green's function, one arrives at the relation
\begin{equation}
\label{eq:condcharged}
\frac{F^2_{\p^\pm} (B)m_{\p}^{2}}{F^2_{\p} m_{\p}^{2}}=\frac{G^{2}_{\p^\pm} (B)}{G^{2}_{\p}}=\frac{\langle\ol{\psi}\psi\rangle_{B}}{\langle\ol{\psi}\psi\rangle_{0}}
,\end{equation}
where we have retained the $m_{\p}^{2}$ in the numerator and denominator to better indicate how this is a generalization of the Gell-Mann-Oakes-Renner relation.
For the neutral pion, $G_{\p^{0}}(B)=G_{\p}$, consistent with the observation $F_{\p^{0}}(B)m_{\p^{0}}^{2}(B)=F_{\p^{0}}m_{\p^{0}}^{2}$.
%

Additionally, 
due to the numerical value of the low-energy constant 
$\overline{l\phantom{l}}$
in
Eq.~\eqref{eq:LEC=1}, 
one has the approximate relation
\begin{equation}
F_{\p^\pm}^{(A2)} 
\approx 
\frac{1}{2} \, F_{\p^\pm}^{(V)}
+ \c O \left(\frac{eB}{\L_\chi^3} \right)
\label{eq:NLOTH}
,\end{equation}
which is satisfied at the 
$10\%$
level given the uncertainty in the extraction of 
$\overline{l\phantom{l}}$
from data. 
This relation appears to be a numerical coincidence, rather than a low-energy theorem.
Nonetheless,
it should be contrasted with the chiral-limit behavior of the results obtained from the Nambu--Jona-Lasinio model study in 
Ref.~\cite[Eq.~(93)]{Coppola:2019uyr}, 
which instead lead to the relation%
\footnote{
Note that we have changed their notation to ours. 
For small magnetic fields, 
their 
$f_{\p^-}^{(V)}$
and
$f_{\p^-}^{(A2)}$
amplitudes each contain an overall factor of 
$eB$, 
while their 
$f_{\p^-}^{(A3)}$
amplitude contains an overall factor of 
$(eB)^2$. 
Removing an overall 
$e B$, 
the latter amplitude accordingly does not appear in 
Eq.~\eqref{eq:relation}, 
having been absorbed into the correction term. 
}
\begin{equation}
F_{\p^\pm}^{(A2)} 
= 
F_{\p^\pm}^{(V)} + \c O \left(\frac{eB}{\L_\chi^3} \right)
\label{eq:relation}
.\end{equation}
This relation is at odds with chiral perturbation theory.

Finally, 
there is a relation between the 
$F^{(A3)}$
amplitude and the anisotropy in the pion dispersion relation.%
\footnote{
While the dispersion relation for charged pions is already anisotropic, 
the form of the charged-pion effective Lagrangian in 
Eq.~\eqref{eq:aniso} 
represents an additional anisotropy beyond the spectrum of 
$D^\m D_\m$
in a fixed background field. 
} 
This was discussed in 
Ref.~\cite{Coppola:2019uyr}, 
but only for the neutral pion. 
In that work, 
the relation is found due to a Goldberger-Treiman relation for the pion coupling to a
constituent quark-antiquark pair in the chiral limit.  
Independent of model assumptions, 
the relation emerges from the structure of operators contained in the
$\c L_6$
chiral Lagrangian. 
Utilizing the result from  
Appendix~\ref{s:L6}, 
the single-pion effective action at 
NNLO 
has the general form
\begin{multline}
\c L
=
\frac{1}{2} \p^0
\left[ - \partial^\m \partial_\m - \frac{(eB)^2}{F_\p^4} c_{\p^0} \partial^\m_\perp \partial_{\perp \m} - m_{\p^0}^2(B)
\right] \p^0
\\
+
\p^+
\left[ - D^\m D_\m - \frac{(eB)^2}{F_\p^4} c_{\p^\pm} D_\perp^\m D_{\perp \m} - m_{\p^\pm}^2(B)
\right]
\p^-
\label{eq:aniso}
,\end{multline}
after wave-function renormalization has been taken into account. 
The general form of this effective action can be argued based on symmetry principles. 
In a uniform magnetic field, 
the Lorentz symmetry group is reduced to an 
$SO(1,1) \times SO(2)$
subgroup%
~\cite{Miransky:2002rp}. 
The effect in chiral perturbation theory can be quantified through the factor
\begin{equation}
Z_\perp^Q
=
1 + \frac{(eB)^2}{F_\p^4} c_{\p^Q}
,\end{equation} 
which represents the residual renormalization factor for the transverse directions.
The deviation from unity must be quite small for 
$e B \sim m_\p^2$; 
but, 
in principle,  
can be deduced from the pion dispersion relation. 
Nonetheless, 
the fact that the same parameter
$c_{\p^Q}$
enters the amplitude 
$F^{(A3)}$
leads to the NNLO relation
\begin{equation}
(eB)^2 
F_{\p^Q}^{(A3)}
=
\left(  
1 - Z_\perp^{Q} 
\right)
F_{\p^Q}^{(A1)}
\label{eq:anisorelation}
,\end{equation}
for both neutral and charged pions.
The result for the neutral pion agrees with that found in 
Ref.~\cite{Coppola:2019uyr}, 
while the charged pion relation was not investigated in their model study.

\section{Pion Decays}
\label{s:decays}

\subsection{Neutral Pion Decay}

\begin{figure}
	\centering 
	\includegraphics[width=0.475\textwidth, angle=0]{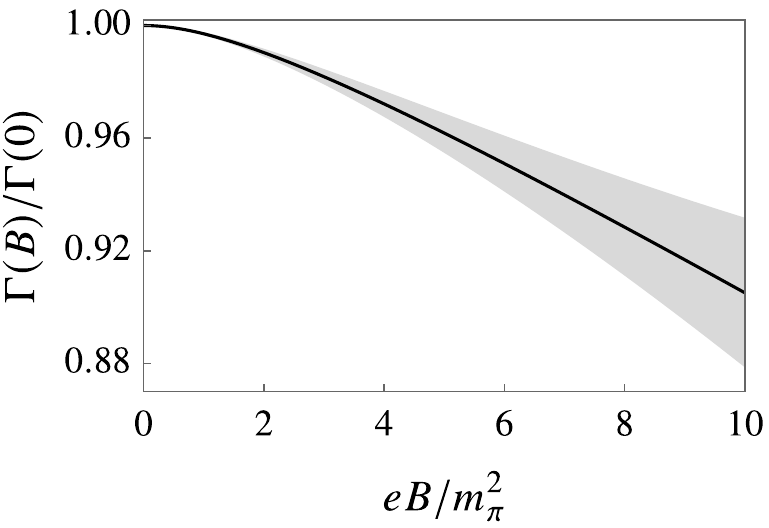}	
	\caption{Neutral pion decay in a magnetic field.
	The ratio of the decay rate Eq.~\eqref{eq:gammagamma} 
	for the anomalous process 
	$\p^0 \to \g \g$ 
	in a magnetic field to that in zero magnetic field
	is plotted as a function of the magnetic field strength. 
	The uncertainty band is generated by varying the relative NNLO correction between
	$\pm (eB)^2 / \L_\chi^4$. } 
	\label{f:neutral}%
\end{figure}

While the weak decay rate of the neutral pion in a magnetic field can be computed from results above, 
electromagnetic processes mediated by the chiral anomaly are considerably more important. 
Decay through the anomalous process 
$\pi^0 \to \g \g$
is the dominant decay mode.
The pertinent term obtained from expanding the two-flavor
WZW
Lagrangian 
Eq.~\eqref{eq:WZW}
is well known
\begin{equation}
\c L_4^\text{WZW}
\supset
- \frac{\a}{\p} \, \frac{\p^0}{F_\p} \b E  \cdot \b B 
\label{eq:WZWpi0}
.\end{equation}
Accounting for all NLO terms in the chiral expansion, 
the two-photon decay rate in a magnetic field becomes
\begin{equation}
\G_{\p^0 \to \g\g}(B)
=
\frac{\a^2 \, m^3_{\p^0}(B)}{4 \p \L_\chi^2 }
\label{eq:gammagamma}
,\end{equation}
and depends on the magnetic field through the neutral pion's magnetic mass.
As shown in 
Fig.~\ref{f:neutral}, 
the decay rate decreases with increasing magnetic field, 
which is entirely driven by the decrease in available energy.

In vanishing magnetic field, 
the anomaly-mediated process
$\p^0 \to e^+ e^-$
is considerably suppressed compared to two-photon decay, 
in part due to two additional electromagnetic vertices
(for a recent review and update, see Ref.~\cite{Husek:2024gnn}). 
The term in 
Eq.~\eqref{eq:WZWpi0}, 
however, 
leads to a new mechanism for this decay in an external magnetic field. 
With the electric field providing a virtual photon,  
the vertex mediates the anomalous process 
\begin{equation}
\pi^0 \overset{B}{\longrightarrow} \, \g^* \longrightarrow e^+ e^-
\label{eq:new}
.\end{equation} 
The coupling between the
$\p^0$ 
and a virtual photon increases linearly with increasing magnetic field. 
For fields satisfying
$e B / m_\p^2 \gg \alpha$, 
the process
$\p^0 \to e^{+} e^{-}$
is dominated by the reaction in 
Eq.~\eqref{eq:new},
as opposed to the  
$B = 0$
mechanism.
Increasing the field strength further, 
the new mechanism becomes a larger fraction of the full decay width. 
To compete with the primary decay mode 
$\p^0 \to \g \g$, 
however, 
one naively requires magnetic fields of size 
$e B / m_\p^2 \sim \a^{-1}$, 
which are considerably beyond the reach of chiral perturbation theory. 
In the regime where our results apply, 
two-photon decay certainly remains the dominant mode for neutral pion decay.

\subsection{Charged Pion Decay}

To compute the weak decay rate of the charged pion, 
we first specify the kinematics. 
For simplicity, 
we describe the calculation for the positively charged pion 
$\p^+ \to \ell^+ \,  \n_\ell$. 
We have verified that the negatively charged pion decay rate is identical, 
which provides a non-trivial check of the final result. 
With the gauge choice employed throughout, 
the good components of momentum for a charged particle are denoted by 
\begin{equation}
\wt P^\m = (E, \b{\wt{P}})
,\quad \text{where} \quad 
\b{\wt{P}} = (P_x, 0, P_z)
.\end{equation}
While 
$\wt P^\m$
is certainly not a four-vector, 
the 
$SO(1,1)$ 
subgroup of longitudinal Lorentz transformations is maintained. 
Note that the interpretation of the 
$P_x$
quantum number as momentum is gauge dependent%
~\cite{THaugset:1993}.

In the 
$N$-th Landau level,  
the charged pion energy is given by
\begin{equation}
E_{N,P_z} = \sqrt{m_{\p^\pm}^2(B) + (2N {+} 1) eB  + P_z^2}
\label{eq:Epion}
,\end{equation} 
where 
the magnetic mass of the charged pion
$m_{\p^\pm}(B)$
appears in 
Eq.~\eqref{eq:mpipm}. 
The spectrum's 
$P_x$--independence reflects the infinite degeneracy of Landau levels. 
The decay rate itself is demonstrably independent of 
$P_x$%
~\cite{Coppola:2018ygv}, 
which provides circumstantial evidence for gauge invariance. 
Direct evidence has been obtained by verifying the total decay rate is identical using 
the symmetric gauge%
~\cite{Coppola:2019wvh}.

The spectrum of an anti-lepton of mass 
$m_\ell$
in the final state can be written as
\begin{equation}
E_{n, p_z}
=
\sqrt{m_\ell^2 + 2 n e B  + p_z^2}
,\end{equation}
where each energy eigenstate has a two-fold infinite degeneracy, 
with degeneracy labels 
$s = \pm \frac{1}{2}$
and 
$p_x$.
The exceptions are states with 
$n = 0$,
for which only the $z$-component of spin
$s = + \frac{1}{2}$
is possible. 
We have used 
$n$
as one of the quantum numbers;
it is related to the usual Landau level quantum number 
$n_L$
through 
$n = n_L - s + \frac{1}{2}$. 
The neutrino final state is labeled by its 
$z$-component of spin 
$s'$
and four-momentum 
$q^\m$.

Treating all momentum quantum numbers of the decay implicitly, 
the coordinate-space amplitude for charged pion leptonic decay is 
\begin{multline}
\c M^{s's}_n(x)
=
G_F V_\text{ud}
\left[ \ol u {}^{s'}_\b{q} (x) \g_\m (1 {-} \g^5) V^s_{n \, \wt{\b p}}(x)
\right] 
\\ 
\times
\big\langle 0  \big| J^{\m +}_L(x) \big| \p^+_{N, \wt {\b P}}\, \big\rangle
\label{eq:M}
,\end{multline}
where 
$G_F$
is the 
Fermi coupling constant, 
and 
$V_\text{ud}$
is the light-quark weak mixing matrix element. 
The coordinate-space spinor 
$\ol u {}^{s'}_\b{q} (x)$
is that for the neutrino, 
which is just the Fourier phase multiplied by the conventionally defined on-shell momentum spinor 
$\ol u{}^{s'}_\b{q} (x) = e^{i q \cdot x} u^{s'}(q)$. 
The coordinate-space spinor for an on-shell anti-lepton is
$V^s_{n \, \wt{\b p}}(x)$, 
and satisfies a generalization of the standard normalization condition
\begin{multline}
\int d^3 x \,
\ol V {}^s_{n \, \wt{\b p}}(x) V^{s'}_{n' \, \wt{\b p}\,'}(x)
\\
= 
- 2 \, m_\ell \, \d^{ss'} \d^{n n'} (2 \p)^2 \d^{(2)} (\wt{\b p} {-} \wt{\b p} \,')
,\end{multline}
when integrated over all space.
Given the left-handed current at NLO in the effective theory, 
one readily computes its single-pion--to--vacuum matrix element
appearing in the amplitude
Eq.~\eqref{eq:M}.

Appealing to conservation of the good components of momentum, 
we define a reduced amplitude through the relation
\begin{multline}
\int d^3 \wt x \, \c M^{s's}_n(x)
\equiv
(2 \p)^3 \d^{(2)} \left( \wt{\b P} {-} \wt{\b p} {-} \wt{\b q} \, \right) 
\\
\times
\d(E_{N,P_z} {-} \, E_{n, p_z} {-} \, E_\b{q})
\,\c M_\text{n, red}^{s's} (y)
,\end{multline}
where the integration is over all coordinates canonically conjugate to the 
good components of momentum. 
We use shorthand for the evaluation at
$x^\m |_{\wt x \, {}^\m = 0}$ 
as simply 
$x = y$. 
The coordinate dependence of the reduced amplitude 
reflects the Fourier transform of the displaced
coordinate overlap between the pion Landau level 
and that of the anti-lepton.

With these specifications, 
the total decay rate for the process
$\p^+ \to \ell^+ \, \n_\ell$
is given by 
\begin{multline}
\G_\ell(B)
=
\frac{1}{2 E_{N,P_z}}
\sum_{n}
\int \frac{d^3 q \, d^2 \wt p}{2 E_\b{q} \, 2 E_{n, p_z} (2\p)^5}
\,
\big\langle \,
|\c M_n|^2 
\big\rangle
\\
\times (2\p)^3 \d^{(2)} \left( \wt{\b P} {-} \wt{\b p} {-} \wt{\b q} \, \right) \, \d(E_{N,P_z} {-} \, E_{n, p_z} {-} \, E_\b{q})
\label{eq:Gamma}
,\end{multline}
where the modulus squared amplitude summed over final state spins is defined as
\begin{equation}
\big\langle \,
|\c M_n|^2 
\big\rangle
=
\sum_{s's}
\Bigg| \int 
dy \, \c M_\text{n, red}^{s's} (y) \, \Bigg|^2
.\end{equation}
Although only 
$n$ 
is labeled, 
$\big\langle \,
|\c M_n|^2 
\big\rangle$
 is additionally a function of the initial-state quantum numbers
$N$ and $P_z$, 
as well as the final-state quantum numbers
$\wt{\b p}$ and $\b q$. 
The momentum 
$P_x$
produces an overall phase in 
$\c M^{s's}_\text{n, red}(y)$
that cancels in the modulus squared.

In the total decay rate, 
the pre-factor 
$E_{N,P_z}^{-1}$
is not invariant under longitudinal Lorentz transformations;
instead, 
a boosted pion with 
$\be = P_z / E_{N,P_z}$
experiences time dilation governed by the factor
 \begin{equation}
 \g = \sqrt{1 - \be^2} = \frac{E_{N,0}}{E_{N,P_z}}
 .\end{equation} 
Accordingly, 
the remaining factors must be invariant under longitudinal Lorentz transformations 
to ensure covariance of the decay rate. 
As the phase space and energy-momentum conserving delta functions in 
Eq.~\eqref{eq:Gamma}
satisfy such invariance, 
it must be that 
$\big\langle \,
|\c M_n|^2 
\big\rangle$
is invariant under the 
$SO(1,1) \times SO(2)$
subgroup of Lorentz transformations,
which we verify below.

Enforcing the decay kinematics through the delta-functions in 
Eq.~\eqref{eq:Gamma}
produces a constraint on the allowed lepton energy levels, 
namely
$n \leq n_\text{max}$, 
where 
\begin{equation}
n_\text{max}
= 
\frac{m^2_{\p^\pm}(B) - m_\ell^2}{2 e B} + N {+} \frac{1}{2}
,\end{equation}
which, 
for convenience, 
is not defined to be an integer. 
Conservation of the good components of momentum fixes the magnitude 
of the neutrino's momentum to be
\begin{multline}
| \b q|
=
\frac{E_{N,P_z} {-} P_z \, x}{1-x^2}
\\
-
\frac{\sqrt{(E_{N,P_z} {-} P_z \, x)^2 - 2 e B (n_\text{max} {-} n) (1 {-} x^2)}}{1-x^2}
,\end{multline}
where
$x = \cos \theta$, 
and 
$\theta$
is the angle of the neutrino's momentum with the respect to the magnetic field. 
The differential decay rate for an initial-state pion in its lowest Landau level can thus be written in the form
\begin{equation}
\frac{d\G_\ell(B)}{dx}
=
\frac{1}{16 \p E_{N,P_z}}
\sum_{n =0}^{\lfloor n_\text{max} \rfloor}
\frac{|\b q| \, \big\langle \, |\c M_n|^2 \big\rangle}
{E_{N,P_z} {-} P_z \, x {-} (1{-}x^2) \,  | \b q |}
,\end{equation}
with the total decay rate given as the integral over all possible angles
$\G_\ell(B) = \int\limits_{-1}^{+1} \frac{d\G_\ell(B)}{dx} dx$.

It remains to evaluate the amplitude squared, 
which is straightforward; 
but, 
the full formula for  
$\big\langle \,
|\c M_n|^2 
\big\rangle$
is rather lengthy and we do not display it
(see Ref.~\cite{Coppola:2018ygv} for the complete expression, 
albeit in slightly different notation).
There is a reduction in the number of terms if we specify that the initial-state pion is in the lowest Landau level 
$N = 0$. 
When the magnetic field is suitably large, 
there will be a large gap between the lowest Landau level and the excited levels. 
In such cases, 
we can assume that all excited states have radiated down to the lowest Landau level before the weak decay. 
The large gap provides a suitably large phase space for the radiative decay, 
which will then occur on shorter time scales than the weak decay.

Restricting to  
$N = 0$, 
there are different contributions to the amplitude for the lepton energy levels
$n = 0$
and 
$n \neq 0$. 
Such differences arise due to the exclusion of the
$s = - \frac{1}{2}$
spin state for 
$n = 0$. 
As the amplitude
$F^{(A3)}$
vanishes at NLO, 
all contributions can be written in terms of three combinations of the remaining amplitudes
\begin{align}
\c A
&= 
F_{\p^\pm}^{(A1)} + e B \, F_{\p^\pm}^{(V)}
,\notag \\
\c B
&=
F_{\p^\pm}^{(A1)} - e B \, F_{\p^\pm}^{(V)}
,\notag \\
\c C
&=
F_{\p^\pm}^{(A1)} + e B \, F_{\p^\pm}^{(A2)}
.\end{align}
In terms of these combinations, 
we have%
\footnote{
Aside from an overall factor of 
$(2 \p)^2$, 
the amplitude squared agrees with the general formula given in 
Ref.~\cite[Eq.~(45)]{Coppola:2018ygv}, 
when specialized to the case of an initial-state pion in the lowest Landau level
$N = 0$.
The difference in normalization is accounted for by a compensating factor of
$(2 \p)^{-2}$
in the decay rate
Eq.~\eqref{eq:Gamma}.
}
\begin{widetext}
\begin{multline}
\big\langle \,
|\c M_n|^2 
\big\rangle
=
8 G_F^2 V_\text{ud}^2 
\, \exp ( - \z )
\Bigg\{
\frac{ \z^n}{n!}
\Bigg[
\frac{\c A^2}{2}
 P^- q^+ P^- p^+
+ 
e B \, \c C^2 \, q^- p^+ 
\frac{(n {-} \z)^2}{\z}
+ 
2 e B \, \c A \c C \, 
P^- p^+
(n - \z)
\Bigg]
\\
+
(1 {-} \d_{n0})
\frac{ \z^{n-1}}{ (n{-}1)!}
\Bigg[
\frac{\c B^2 }{2}
P^+ q^- P^+ p^-
- 
2 
e B \, \c A \c B \,
P^+ P^-
\, \z
- 
2 
e B \, \c B \c C \, 
P^+ q^-
(n {-} \z)
\Bigg]
\Bigg\}
\label{eq:SumM}
,\end{multline}
\end{widetext}
where we employ the abbreviation 
\begin{equation}
\z
=
\frac{\b q_\perp^2}{2 e B} 
,\end{equation}
with 
$\b q_\perp$
as the neutrino momentum transverse to the magnetic field.
The dependence on 
$\z$
ensures that the decay rate is 
$SO(2)$
invariant.  
For any 
$(1{+}1)$-dimensional 
Lorentz vector with components 
$a^0$
and
$a^3$,
the corresponding light-front components employed above are defined by 
\begin{equation}
a^\pm = a^0 \pm a^3
.\end{equation}
Products of the form 
$a^+ b^-$
and 
$a^- b^+$
are invariant under 
longitudinal Lorentz boosts. 
Accordingly, 
the modulus squared amplitude in 
Eq.~\eqref{eq:SumM}
is manifestly invariant under 
$SO(1,1)$.  
This confirms that time dilation is the only effect on the total decay rate of a longitudinally boosted pion 
compared to a pion at rest 
$P_z = 0$.

\begin{figure}
	\centering 
	\includegraphics[width=0.475\textwidth, angle=0]{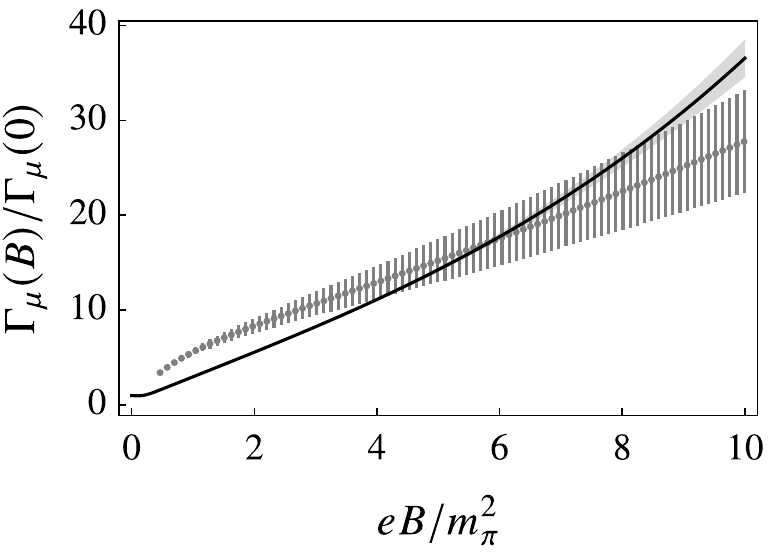}
	\vskip0.75em
	\includegraphics[width=0.475\textwidth, angle=0]{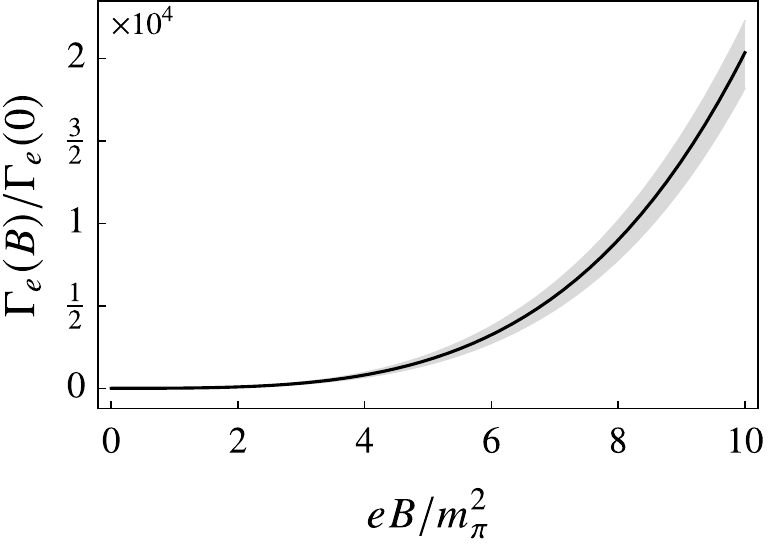}	
\caption{Charged pion decay in a magnetic field.
	The ratio of the decay rate Eq.~\eqref{eq:Gamma} 
	in a magnetic field to that in zero magnetic field Eq.~\eqref{eq:Gamma0}
	is plotted as a function of the magnetic field strength
	for  muonic and electronic decay modes. 
	The uncertainty band is generated by varying the relative NNLO correction between
	$\pm 2 (eB)^2 / \L_\chi^4$, 
	as well as varying the low-energy constant 
	$\overline{l\phantom{l}}$
	within its uncertainties. Staggered quark lattice data is plotted for the muonic decay width.} 
	\label{f:charged}%
\end{figure}

\begin{figure}
	\centering 
	\includegraphics[width=0.475\textwidth, angle=0]{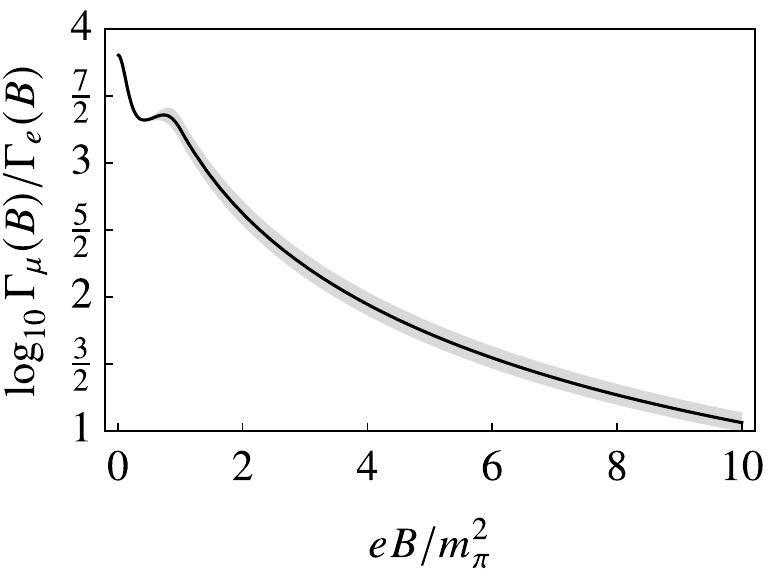}
\caption{Comparison of the charged pion's muonic and electronic decay modes. 
	The logarithm of the ratio of partial widths (muonic to electronic) is plotted as a function of the magnetic field. 
	While we include an uncertainty band generated in the same manner as  
	Fig.~\ref{f:charged}, 
	it can barely be discerned from the width of the curve plotted.} 
	\label{f:ratio}%
\end{figure}

The charged pion decay rate for muonic and electronic decay modes is 
plotted as a function of the magnetic field strength in 
Fig.~\ref{f:charged}. 
Each decay rate is plotted as a ratio to the corresponding rate in vanishing magnetic field
\begin{equation}
\G_\ell(0)
=
\frac{G_F^2 V_\text{ud}^2 F_\p^2}{4 \p} m_\p m^2_\ell \left( 1 - \frac{m_\ell^2}{m_\p^2} \right)^2
\label{eq:Gamma0}
.\end{equation}
This rate, 
however, 
does not directly arise from the zero-field limit of 
Eq.~\eqref{eq:Gamma}, 
due to the restriction of the initial-state pion to its lowest Landau level.%
\footnote{
It has been shown, 
however, 
that the zero-field result properly emerges after adjusting the overall normalization 
from that of a discrete Landau level to that of a continuum zero-momentum state%
~\cite{Coppola:2018ygv}.
}  
Both partial widths have increased compared to their corresponding zero-field widths. 
In the case of the electronic decay mode, 
the width is substantially greater, 
because the magnetic field allows the electron to overcome helicity suppression. 
The electronic decay mode, 
however, 
becomes comparable to the muonic one as the magnetic field strength increases.
In zero magnetic field, 
the ratio of partial widths is
$\G_\m(0)/ \G_e(0) \approx 10^4$. 
In Fig.~\ref{f:ratio}, 
the magnetic field dependence of this ratio is plotted as a function of  
$e B / m_\p^2$. 
While the muonic decay mode remains more probable than the electronic decay mode, 
it is no longer overwhelmingly preferred. 
The ratio approaches 
$10.1(5)$
for the largest magnetic field shown. 
The observed bump in the ratio for small values of the magnetic field is a consequence of the muonic decay rate rising rapidly while the electronic decay rate remains flat.
We have also plotted lattice data for the muonic decay width performed in the lowest landau approximation in Fig.~\ref{f:charged}. Unsurprisingly, the approximation breaks down for weak fields and produces agreement only for larger magnetic fields. The disagreement for smaller fields, $eB\sim 0-5m_{\p}^{2}$, is a consequence not only of the lowest Landau approximation but also a disagreement in the pion decay constants.

\begin{figure}
	\centering 
	\includegraphics[width=0.475\textwidth, angle=0]{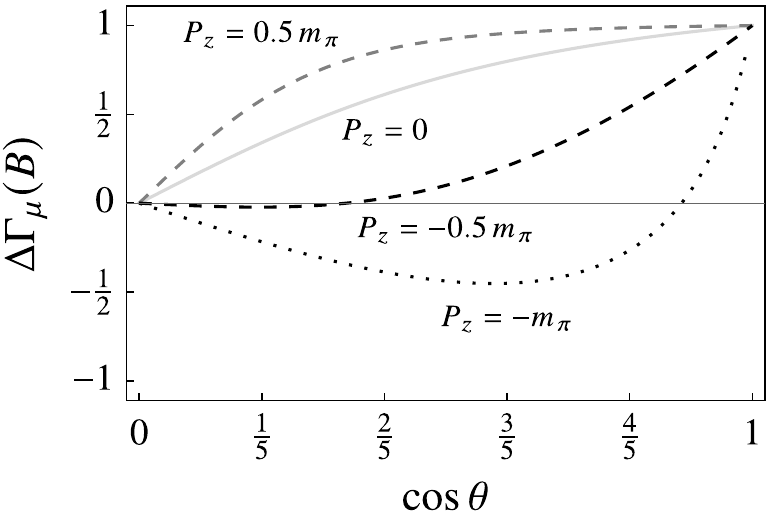}
\caption{Angular asymmetry in the differential decay width.
Using a fixed value of the magnetic field, 
the asymmetry 
Eq.~\eqref{eq:asym} 
of the decay 
$\p^+ \to \m^+ \n_\m$ 
is plotted as a function of 
$x = \cos \theta$, 
for a few values of the pion's longitudinal momentum 
$P_z$. 
} 
	\label{f:asym}%
\end{figure}

While the total decay rate only depends on 
$P_z$
through the time-dilation factor~\footnote{Since decay width has units of inverse time, one expects the effects of time dilation to be inherent in it. In the zero-field case, the effects of time dilation are entirely encoded in the factor of $\frac{1}{E_{\mathbf{P}}}$ with the phase space integral involving the Feynman amplitude being Lorentz invariant as the measure is Lorentz invariant and the amplitude a function of Lorentz invariant momenta, $P\cdot p$, $P\cdot q$ and $p\cdot q$. In the presence of a magnetic field, we find that time dilation effects are entirely encoded in the overall factor of $\frac{1}{E_{0,P_{z}}}$ as a numerical evaluation of the ratio for example cases confirm
\begin{align}
\frac{\Gamma(P_{z})/\Gamma(P_{z}=0)}{E_{0,0}/E_{0,P_{z}}}=1\ .
\end{align}
}, 
the differential decay rate has a more complicated dependence on the 
pion's longitudinal momentum. 
To investigate this dependence, 
we form the angular asymmetry in the differential decay rate
\begin{equation}
\D \G_\ell(B)
=
\frac{\frac{d\G_\ell(B)}{dx} \big|_x - \frac{d\G_\ell(B)}{dx} \big|_{-x}}
{\frac{d\G_\ell(B)}{dx} \big|_x + \frac{d\G_\ell(B)}{dx} \big|_{-x}}
\label{eq:asym}
.\end{equation}
The component of the neutrino's momentum in the direction of the magnetic field can be positive or negative. 
When it is equally likely to be positive or negative, 
the asymmetry vanishes
$\D \G_\ell(B) = 0$. 
The angular asymmetry is shown in 
Fig.~\ref{f:asym}
for the muonic decay mode at a fixed value of the external field. 
Already for a pion at rest
$P_z = 0$, 
the neutrino emerges with a $z$-component of momentum preferentially in the direction of the magnetic field. 
Of course, 
this bias for the magnetic field direction only increases when the pion is also moving in the same direction. 
When the pion is moving in the direction opposite of the magnetic field 
with a suitably large momentum,
however,
there can be more neutrinos emerging with their $z$-component of momentum anti-aligned with the field.
This can happen for a range of 
$\cos \theta$,
away from the value
$\cos \theta = 1$.

\section{Summary and conclusions}
\label{s:sum}

The hadronic left-handed vector current in an external magnetic field is studied 
systematically using chiral perturbation theory. 
This analysis provides model-independent constraints on pion matrix elements 
that arise due to the pattern of explicit and spontaneous symmetry breaking in low-energy QCD. 
These constraints constitute low-energy theorems that must be satisfied by 
lattice QCD calculations, and additionally provide stringent tests for models of hadronic physics
(for example, 
the relation 
Eq.~\eqref{eq:NLOTH} 
is not satisfied by the results of the model study in 
Ref.~\cite{Coppola:2019uyr}). 
While the applicability of chiral perturbation theory limits one to perturbatively small fields 
$e B \ll \L_\chi^2$, 
the results serve to anchor the behavior of observables in the regime 
$e B \sim m_\p^2$.

As the chiral anomaly is the source of the vector-pseudoscalar coupling in an external magnetic field, 
the 
$F^{(V)}$
amplitude provides a relatively straightforward means to directly measure the chiral anomaly in lattice QCD. 
Beyond the pioneering work in 
Ref.~\cite{Bali:2018sey}, 
further calculations are needed to address the tension 
Eq.~\eqref{eq:anomaly}
between the result for staggered quarks at the physical quark masses
and the chiral anomaly. 
Additional calculations are warranted for the 
$F^{(A1)}$
amplitude in a magnetic field. 
The small-field behavior observed empirically from lattice QCD data
is at odds with that obtained from chiral perturbation theory
Eq.~\eqref{eq:FA1small}. Additionally, predictions of the Nambu-Jona-Lasinio model are at odds with chiral perturbation theory as illustrated in Fig.~\ref{f:FA1NJL}. Disagreement is additionally observed in the magnetic polarizability as discussed in Section~\ref{sec:NJL}. Notably disagreement is strongest for weak fields for which the relevant degrees of freedom from an effective field theory pespective are pions, not quarks.
The Gell-Mann--Oakes--Renner relations in a magnetic field
Eqs.~\eqref{eq:cond} and \eqref{eq:condcharged}, 
furthermore,
allow one to relate magnetic catalysis to the increase in
$F^{(A1)}$
amplitudes in a magnetic field.

The remaining axial-vector amplitudes, 
$F^{(A2)}$
and
$F^{(A3)}$, 
provide an opportunity to extract low-energy constants of chiral perturbation theory from lattice QCD. 
For both of these amplitudes, 
the same low-energy constants enter the pion spectrum, 
which can thereby be utilized for a consistency check on the calculations, 
or 
provide additional data to constrain these numerically small contributions. 
For 
$F^{(A2)}$, 
the underlying physics is that of the charged pion magnetic polarizability; 
whereas, 
for 
$F^{(A3)}$, 
the underlying physics is the magnetic field induced anisotropy in the pion dispersion relation.

With the model-independent results from chiral perturbation theory, 
the dominant decays of neutral and charged pions are computed in the 
$e B \sim m_\p^2$
regime. 
For the neutral pion, 
the electromagnetic decay rate via the anomalous process 
$\p^0 \to \g \g$
appears in Fig.~\ref{f:neutral}, 
where a modest decrease with increasing magnetic field is shown. 
Although other decay mechanisms become available in an external magnetic field, 
the zero-field decay mode remains the dominant one. 
For the charged pion, 
we explicitly verify the 
$SO(1,1) \times SO(2)$
covariance of its weak decay rate. 
Time-dilation is thereby the only effect on the total decay rate of a longitudinally moving pion. 
The electronic decay mode is greatly enhanced, 
consistent with the lack of helicity suppression in a magnetic field%
~\cite{Bali:2018sey,Coppola:2019idh}. 
In the regime where our results apply, 
the muonic decay mode is still dominant, 
but only by at most a factor of 
$\sim 10$. 
For the differential decay rate, 
we explore the longitudinal momentum dependence through the angular asymmetry 
Eq.~\eqref{eq:asym}. 
Rather large momentum antiparallel to the field 
(for example, 
$P_z {=} - m_\p$)
is needed to overcome the bias for the magnetic field direction 
in the emerging neutrinos.

The analysis can be generalized to three-flavor chiral perturbation theory, 
as well as carried out at non-zero temperature. 
Higher-order effects would be interesting to ascertain, 
even if only their signs. 
The behavior of the magnetic mass of the neutral pion, 
for example, 
is intriguing, 
for which the natural question is whether the NNLO contribution also decreases with increasing magnetic field. 
Additionally the qualitative behavior of the NNLO contribution to the axial-vector amplitude
$F^{(A1)}$
may shed light on the strikingly different behavior seen in lattice QCD data. 
A higher-order calculation of the chiral anomaly in an external magnetic field would 
be beneficial to address the low-energy QCD prediction for the 
$F^{(V)}$
amplitude, 
and its subsequent extraction from future lattice QCD data. Finally, a study of finite volume corrections of pion decay constants in a magnetic background may prove fruitful in understanding the large discrepancy between lattice QCD and chiral perturbation theory~\cite{Adhikari:2023fdl}.

\section*{Acknowledgements}
We thank Igor Shovkovy for helpful comments on the manuscript and the authors of Ref.~\cite{Bali:2018sey} for sharing lattice data. 
P.A. is supported in part by the U.S. Department of Energy, Office of Science, Office of Nuclear Physics and Quantum Horizons Program under Award Number 
DE-SC0024385. 
P.A. acknowledges the hospitality of The City College of New York, the Graduate Center of the City University of New York, and that of the Kavli Institute for Theoretical Physics, Santa Barbara, 
through which the research was supported in part by the
U.S. National Science Foundation under 
Grant No. 
NSF PHY-1748958. 

\appendix

\section{Tree-Level Contribution at Next-To-Next-To-Leading Order}
\label{s:L6}

The first non-vanishing contribution to the 
$F^{(A3)}$
amplitude in 
Eq.~\eqref{eq:JA}
occurs at NNLO in chiral perturbation theory from two-loop and one-loop diagrams, 
as well as from tree-level operators. 
The full NNLO calculation is beyond the scope of this work; 
however, 
the structure of the answer, 
aside from chiral logarithms,  
can be deduced solely from the tree-level contribution. 
To this end, 
we require terms from the two-flavor
$\c L_6$
Lagrangian%
~\cite{Bijnens:1999sh}
\begin{eqnarray}
Y_{31}
&=&
\langle \hat{f}_{+ \m \n} \hat{f}_+^{\m \a}  u^\n u_\a \rangle
,\notag\\
Y_{32}
&=&
\langle \hat{f}_{+ \m \n} \hat{f}_+^{\m \a} u_\a  u^\n \rangle
,\notag\\
Y_{33}
&=&
\Big\langle \hat{f}_{+ \m \n} \left( u_\a \hat{f}_+^{\m \a}  u^\n  +  u^\n \hat{f}_+^{\m \a}  u_\a  \right) \Big\rangle
\label{eq:L6}
.\end{eqnarray}
These operators have been written using the traceless vielbein field
\begin{equation}
u_\m 
= 
i \left[
u^\dagger \left( \partial_\m - i \hat{R}_\m \right) u
-
u \left( \partial_\m - i \hat{L}_\m \right) u^\dagger
\right]
\label{eq:umu}
,\end{equation}
and the chirally covariant field-strength tensor
$\hat{f}_+^{\m \n}$, 
which is defined by 
\begin{equation}
\hat{f}_+^{\m \n}
=
u \, \hat{L}^{\m \n} u^\dagger + u^\dagger \hat{R}^{\m \n} u
.\end{equation}
The compensator field 
$u$
is determined from the coset field
$U$
appearing in 
Eq.~\eqref{eq:L2}
through the relation
$u^2 = U$. 
The Lagrangian is formed from sums of the
$Y_j$ 
operators weighted by 
$F^{-2} c_j$, 
so that the corresponding low-energy constants 
$c_j$
are dimensionless.

The construction of the minimal operator basis in 
Ref.~\cite{Bijnens:1999sh}
makes explicit use of traceless left- and right-handed vector fields. 
For an electromagnetic field, 
there is no effect on 
$u_\m$
in 
Eq.~\eqref{eq:umu}, 
because the isoscalar component of the gauge field decouples. 
Similarly, 
isoscalar gauge fields decouple from the terms of the 
$\c L_4$
Lagrangian density 
Eq.~\eqref{eq:L4}, 
aside from a pion-independent contact term. 
The same is not true for 
$\c L_6$, 
and two additional operators with isoscalar gauge fields are needed. 
We take these operators to be
\begin{eqnarray}
\wt Y_{31}
&=&
\langle L_{\m \n} {+} R_{\m \n} \rangle
\langle L^{\m \a} {+} R^{\m \a} \rangle
\langle u^\n u_\a \rangle
,\notag \\
\wt Y_{32}
&=&
i \langle L_{\m \n} {+} R_{\m \n} \rangle
\langle \hat{f}_{+}^{\m \a} \left[ u^\n, u_\a \right] \rangle
\label{eq:L6tilde}
.\end{eqnarray}
Note that an operator similar to 
$\wt Y_{32}$
but with 
$i \left[ u^\n, u_\a \right]$
replaced by 
$\left\{ u^\n, u_\a \right\}$
identically vanishes. 
The isoscalar gauge field contributions to the Lagrangian require a sum over the 
$\wt Y_j$
operators weighted by 
$F^{-2} \, \wt c_j$, 
where the corresponding low-energy constants 
$\wt c_j$
are dimensionless.

With the requisite terms of 
$\c L_6$
spelled out, 
we can obtain the isovector left-handed current in a background electromagnetic field using 
Eq.~\eqref{eq:JL}. 
This produces the contribution
\begin{equation}
\left( J_L^{\m Q} \right)_\text{NNLO}
=
- c_{\p^Q} \frac{e^2}{F^3}
F^{\m \a} F_{\m \be} \, D^\be \p^Q
,\end{equation} 
from which the axial-vector amplitude 
$F_{\p^Q}^{(A3)}$
reported in 
Eq.~\eqref{eq:FA3}
follows. 
The parameters 
$c_{\p^Q}$
involve linear combinations of the low-energy constants
\begin{equation}
c_{\p^Q}
=
4 \left[ c_{31} + c_{32} + 2 \big( 1{-} 2 | Q| \big) c_{33} + \frac{4}{9} \,  \wt c_{31} \right]
+
f_{\p^Q}(m_\p)
\label{eq:cs}
,\end{equation}
for which the contribution from the
$\wt Y_{32}$
operator drops out. 
The running of 
$c_{33}$
obtained in 
Ref.~\cite{Bijnens:1999hw}, 
for example, 
points to the necessity of chiral logarithms from loop diagrams, 
which have been denoted by 
$f_{\p^Q}(m_\p)$. 
Accordingly, 
the sum of these contributions renders the
$c_{\p^Q}$
renormalization scale independent.

An additional feature of the operators listed in 
Eqs.~\eqref{eq:L6} 
and 
\eqref{eq:L6tilde}
is that they modify the pion dispersion relation
in an anisotropic way. 
Expanding these terms of the NNLO chiral Lagrangian density in a uniform magnetic field, 
we have the two-pion terms
\begin{equation}
\c L_6
\supset
\frac{(eB)^2}{F^4}
\Bigg[
\frac{1}{2} c_{\p^0}
\, \partial_\perp^\m \p^0 \partial_{\perp \m} \p^0
+
c_{\p^\pm} 
D_\perp^\m \p^+ D_{\perp \m} \p^-
\Bigg]
\label{eq:L6mod}
,\end{equation}
where 
$a^\m_\perp = (0, a^1, a^2, 0)$
is used to denote the components of a four-vector that are transverse to the time and field directions. 
The parameters
$c_{\p^Q}$
are exactly the same linear combinations of renormalized low-energy constants and chiral logarithms appearing in 
Eq.~\eqref{eq:cs}. 
As a result, 
the anisotropy in the dispersion relations of neutral and charged pions is related to the 
$F^{(A3)}$
amplitude, 
which is exhibited in
Eq.~\eqref{eq:anisorelation}. 
\section{Pseudoscalar Current and Renormalization}
\label{s:JP}
The pseudoscalar current follows from its definition, Eq.~\eqref{eq:def-JP}. The relevant contribution arises through the leading chiral Lagrangian and the NLO terms proportional to $l_{3}$ and $l_{4}$
\begin{align}
\label{eq:def-JP}
\hspace{-10pt}
J_{P}^{Q}&=2B_{0}F\p^{Q}-\frac{B_{0}}{3F}\left[(\p^{0})^{2}+2\p^{+}\p^{-}\right]\p^{Q}+J^{Q}_{P,\rm ct}\\
\hspace{-10pt}
J^{Q}_{P, \textrm{ct}}&=\frac{4B_{0}m^{2}}{F}(l_{3}+l_{4})\pi^{Q}
.\end{align}
The counterterm contribution, $J_{P, \textrm{ct}}$, resides in the next-to-leading order Lagrangian
\begin{align}
\nonumber
\c L_{4}\supset& \frac{1}{16}(h_{1}-h_{3}+l_{3})\langle\chi U^{\dagger}+U\chi^{\dagger}\rangle^{2}\\
\nonumber
-&\frac{1}{8}(h_{1}-h_{3}-l_{4})\langle\,\chi U^{\dagger}\chi U^{\dagger}+U\chi ^{\dagger}U\chi ^{\dagger}\,\rangle\ .
\end{align}
A further ingredient is the wavefunction renormalization factor for the pions,
\begin{align}
\sqrt{Z_{\p^{\pm}}}&=1+\frac{1}{6F^{2}}(g_0+g_{B})-\frac{l_{4}m^{2}}{F^{2}}\\
\sqrt{Z_{\p^{0}}}&=1+\frac{1}{3F^{2}}g_{B}-\frac{l_{4}m^{2}}{F^{2}}\ .
\end{align}
$g_{0}$ is the divergent and electromagnetically neutral tadpole contribution or equivalently the coincident Green's function while {$\wt{g}_{B}=-\frac{eB}{(4\p)^{2}}\c{I}(\frac{m^{2}}{eB})$} is the pure magnetic field contribution to the coincident Green's function, $g_{B}=g_{0}+\wt{g}_{B}$ for the charged tadpole. The magnetic field does not introduce further divergences in the tadpole. The renormalized pseudoscalar current follows upon performing a wavefunction renormalization, $\pi^{Q}\rightarrow \sqrt{Z_{\p^{Q}}}\,\pi^{Q}$. 
$G_{\p^{Q}}(B)$ is presented in Eq.~\ref{eq:GB}, and reduces to the zero field result of Ref.~\cite{Gasser:1983yg}.
\bibliography{bibly}

@article{Bijnens:1999hw,
    author = "Bijnens, J. and Colangelo, G. and Ecker, G.",
    title = "{Renormalization of chiral perturbation theory to order $p^6$}",
    eprint = "hep-ph/9907333",
    archivePrefix = "arXiv",
    reportNumber = "LU-TP-99-14, ZU-TH-18-99, UWTHPH-1999-42",
    doi = "10.1006/aphy.1999.5982",
    journal = "Annals Phys.",
    volume = "280",
    pages = "100--139",
    year = "2000"
}

@article{Coppola:2018vkw,
    author = "Coppola, M. and G\'omez Dumm, D. and Scoccola, N. N.",
    title = "{Charged pion masses under strong magnetic fields in the NJL model}",
    eprint = "1802.08041",
    archivePrefix = "arXiv",
    primaryClass = "hep-ph",
    doi = "10.1016/j.physletb.2018.04.043",
    journal = "Phys. Lett. B",
    volume = "782",
    pages = "155--161",
    year = "2018"
}

@article{Fayazbakhsh:2012vr,
    author = "Fayazbakhsh, Sh. and Sadeghian, S. and Sadooghi, N.",
    title = "{Properties of neutral mesons in a hot and magnetized quark matter}",
    eprint = "1206.6051",
    archivePrefix = "arXiv",
    primaryClass = "hep-ph",
    doi = "10.1103/PhysRevD.86.085042",
    journal = "Phys. Rev. D",
    volume = "86",
    pages = "085042",
    year = "2012"
}

@article{Kamikado:2013pya,
    author = "Kamikado, Kazuhiko and Kanazawa, Takuya",
    title = "{Chiral dynamics in a magnetic field from the functional renormalization group}",
    eprint = "1312.3124",
    archivePrefix = "arXiv",
    primaryClass = "hep-ph",
    doi = "10.1007/JHEP03(2014)009",
    journal = "Journal of High Energy Physics",
    volume = "03",
    pages = "009",
    year = "2014"
}

@article{Avancini:2016fgq,
    author = "Avancini, Sidney S. and Farias, Ricardo L. S. and Benghi Pinto, Marcus and Tavares, William R. and Tim\'oteo, Varese S.",
    title = "{$\pi_0$ pole mass calculation in a strong magnetic field and lattice constraints}",
    eprint = "1606.05754",
    archivePrefix = "arXiv",
    primaryClass = "hep-ph",
    doi = "10.1016/j.physletb.2017.02.002",
    journal = "Phys. Lett. B",
    volume = "767",
    pages = "247--252",
    year = "2017"
}

@article{Fayazbakhsh:2013cha,
    author = "Fayazbakhsh, Sh. and Sadooghi, N.",
    title = "{Weak decay constant of neutral pions in a hot and magnetized quark matter}",
    eprint = "1306.2098",
    archivePrefix = "arXiv",
    primaryClass = "hep-ph",
    doi = "10.1103/PhysRevD.88.065030",
    journal = "Phys. Rev. D",
    volume = "88",
    pages = "065030",
    year = "2013"
}

@article{GomezDumm:2017jij,
    author = "G\'omez Dumm, D. and Izzo Villafa\~ne, M. F. and Scoccola, N. N.",
    title = "{Neutral meson properties under an external magnetic field in nonlocal chiral quark models}",
    eprint = "1710.08950",
    archivePrefix = "arXiv",
    primaryClass = "hep-ph",
    doi = "10.1103/PhysRevD.97.034025",
    journal = "Phys. Rev. D",
    volume = "97",
    pages = "034025",
    year = "2018"
}

@article{Simonov:2015xta,
    author = "Simonov, Yu. A.",
    title = "{Pion decay constants in a strong magnetic field}",
    eprint = "1503.06616",
    archivePrefix = "arXiv",
    primaryClass = "hep-ph",
    doi = "10.1134/S1063778816030170",
    journal = "Phys. Atom. Nucl.",
    volume = "79",
    pages = "455--460",
    year = "2016"
}

@article{Andreichikov:2018wrc,
    author = "Andreichikov, M. A. and Simonov, Yu. A.",
    title = "{Chiral physics in the magnetic field with quark confinement contribution}",
    eprint = "1805.11896",
    archivePrefix = "arXiv",
    primaryClass = "hep-ph",
    doi = "10.1140/epjc/s10052-018-6384-x",
    journal = "Eur. Phys. J. C",
    volume = "78",
    pages = "902",
    year = "2018"
}

@article{Hofmann:2020dvz,
    author = "Hofmann, Christoph P.",
    title = "{Pion Pressure in a Magnetic Field}",
    eprint = "2004.01247",
    archivePrefix = "arXiv",
    primaryClass = "hep-ph",
    doi = "10.1103/PhysRevD.101.114031",
    journal = "Phys. Rev. D",
    volume = "101",
    pages = "114031",
    year = "2020"
}

@article{Hofmann:2020ism,
    author = "Hofmann, Christoph P.",
    title = "{Chiral Perturbation Theory Analysis of the Quark Condensate in a Magnetic Field}",
    eprint = "2006.07717",
    archivePrefix = "arXiv",
    primaryClass = "hep-ph",
    doi = "10.1103/PhysRevD.102.094010",
    journal = "Phys. Rev. D",
    volume = "102",
    pages = "094010",
    year = "2020"
}

@article{Hofmann:2020lfp,
    author = "Hofmann, Christoph P.",
    title = "{Thermomagnetic properties of QCD}",
    eprint = "2012.06461",
    archivePrefix = "arXiv",
    primaryClass = "hep-ph",
    doi = "10.1103/PhysRevD.104.014025",
    journal = "Phys. Rev. D",
    volume = "104",
    pages = "014025",
    year = "2021"
}

@article{Zhang:2002xv,
    author = "Zhang, Bing and Dai, Z. G. and Meszaros, P.",
    title = "{High-energy neutrinos from magnetars}",
    eprint = "astro-ph/0210382",
    archivePrefix = "arXiv",
    doi = "10.1086/377192",
    journal = "Astrophys. J.",
    volume = "595",
    pages = "346--351",
    year = "2003"
}

@article{Harding:2006qn,
    author = "Harding, Alice K. and Lai, Dong",
    title = "{Physics of Strongly Magnetized Neutron Stars}",
    eprint = "astro-ph/0606674",
    archivePrefix = "arXiv",
    doi = "10.1088/0034-4885/69/9/R03",
    journal = "Rept. Prog. Phys.",
    volume = "69",
    pages = "2631",
    year = "2006"
}

@article{Vachaspati:1991nm,
    author = "Vachaspati, T.",
    title = "{Magnetic fields from cosmological phase transitions}",
    doi = "10.1016/0370-2693(91)90051-Q",
    journal = "Phys. Lett. B",
    volume = "265",
    pages = "258--261",
    year = "1991"
}

@article{Vachaspati:2020blt,
    author = "Vachaspati, Tanmay",
    title = "{Progress on cosmological magnetic fields}",
    eprint = "2010.10525",
    archivePrefix = "arXiv",
    primaryClass = "astro-ph.CO",
    doi = "10.1088/1361-6633/ac03a9",
    journal = "Rept. Prog. Phys.",
    volume = "84",
    pages = "074901",
    year = "2021"
}

@article{Kharzeev:2007jp,
    author = "Kharzeev, Dmitri E. and McLerran, Larry D. and Warringa, Harmen J.",
    title = "{The Effects of topological charge change in heavy ion collisions: Event by event P and CP violation}",
    eprint = "0711.0950",
    archivePrefix = "arXiv",
    primaryClass = "hep-ph",
    doi = "10.1016/j.nuclphysa.2008.02.298",
    journal = "Nucl. Phys. A",
    volume = "803",
    pages = "227--253",
    year = "2008"
}

@article{Skokov:2009qp,
    author = "Skokov, V. and Illarionov, A. Yu. and Toneev, V.",
    title = "{Estimate of the magnetic field strength in heavy-ion collisions}",
    eprint = "0907.1396",
    archivePrefix = "arXiv",
    primaryClass = "nucl-th",
    doi = "10.1142/S0217751X09047570",
    journal = "Int. J. Mod. Phys. A",
    volume = "24",
    pages = "5925--5932",
    year = "2009"
}

@article{Deng:2012pc,
    author = "Deng, Wei-Tian and Huang, Xu-Guang",
    title = "{Event-by-event generation of electromagnetic fields in heavy-ion collisions}",
    eprint = "1201.5108",
    archivePrefix = "arXiv",
    primaryClass = "nucl-th",
    doi = "10.1103/PhysRevC.85.044907",
    journal = "Phys. Rev. C",
    volume = "85",
    pages = "044907",
    year = "2012"
}

@article{Inghirami:2019mkc,
    author = "Inghirami, Gabriele and Mace, Mark and Hirono, Yuji and Del Zanna, Luca and Kharzeev, Dmitri E. and Bleicher, Marcus",
    title = "{Magnetic fields in heavy ion collisions: flow and charge transport}",
    eprint = "1908.07605",
    archivePrefix = "arXiv",
    primaryClass = "hep-ph",
    doi = "10.1140/epjc/s10052-020-7847-4",
    journal = "Eur. Phys. J. C",
    volume = "80",
    pages = "293",
    year = "2020"
}

@article{PhysRevX.14.011028,
  title = {Observation of the Electromagnetic Field Effect via Charge-Dependent Directed Flow in Heavy-Ion Collisions at the Relativistic Heavy Ion Collider},
  author = {Abdulhamid, M. I. and Aboona, B. E. and Adam, J. and Adams, J. R. and Agakishiev, G. and Aggarwal, I. and Aggarwal, M. M. and Ahammed, Z. and Aitbaev, A. and others},
  collaboration = {STAR Collaboration},
  journal = {Phys. Rev. X},
  volume = {14},
  pages = {011028},
  numpages = {12},
  year = {2024},
  month = {Feb},
  publisher = {American Physical Society},
  doi = {10.1103/PhysRevX.14.011028}
  }

@article{DElia:2010abb,
    author = "D'Elia, Massimo and Mukherjee, Swagato and Sanfilippo, Francesco",
    title = "{QCD Phase Transition in a Strong Magnetic Background}",
    eprint = "1005.5365",
    archivePrefix = "arXiv",
    primaryClass = "hep-lat",
    doi = "10.1103/PhysRevD.82.051501",
    journal = "Phys. Rev. D",
    volume = "82",
    pages = "051501",
    year = "2010"
}

@article{DElia:2011koc,
    author = "D'Elia, Massimo and Negro, Francesco",
    title = "{Chiral Properties of Strong Interactions in a Magnetic Background}",
    eprint = "1103.2080",
    archivePrefix = "arXiv",
    primaryClass = "hep-lat",
    doi = "10.1103/PhysRevD.83.114028",
    journal = "Phys. Rev. D",
    volume = "83",
    pages = "114028",
    year = "2011"
}

@article{Bonati:2013vba,
    author = "Bonati, Claudio and D'Elia, Massimo and Mariti, Marco and Negro, Francesco and Sanfilippo, Francesco",
    title = "{Magnetic susceptibility and equation of state of $N_f=2+1$ QCD with physical quark masses}",
    eprint = "1310.8656",
    archivePrefix = "arXiv",
    primaryClass = "hep-lat",
    reportNumber = "IFUP-TH-2013-22, LPT-13-82",
    doi = "10.1103/PhysRevD.89.054506",
    journal = "Phys. Rev. D",
    volume = "89",
    pages = "054506",
    year = "2014"
}

@article{Bonati:2013lca,
    author = "Bonati, Claudio and D'Elia, Massimo and Mariti, Marco and Negro, Francesco and Sanfilippo, Francesco",
    title = "{Magnetic Susceptibility of Strongly Interacting Matter across the Deconfinement Transition}",
    eprint = "1307.8063",
    archivePrefix = "arXiv",
    primaryClass = "hep-lat",
    reportNumber = "IFUP-TH-2013-15",
    doi = "10.1103/PhysRevLett.111.182001",
    journal = "Phys. Rev. Lett.",
    volume = "111",
    pages = "182001",
    year = "2013"
}

@article{Bali:2011qj,
    author = "Bali, G. S. and Bruckmann, F. and Endrodi, G. and Fodor, Z. and Katz, S. D. and Krieg, S. and Schafer, A. and Szabo, K. K.",
    title = "{The QCD phase diagram for external magnetic fields}",
    eprint = "1111.4956",
    archivePrefix = "arXiv",
    primaryClass = "hep-lat",
    doi = "10.1007/JHEP02(2012)044",
    journal = "Journal of High Energy Physics",
    volume = "02",
    pages = "044",
    year = "2012"
}

@article{Bali:2012zg,
    author = "Bali, G. S. and Bruckmann, F. and Endrodi, G. and Fodor, Z. and Katz, S. D. and Schafer, A.",
    title = "{QCD quark condensate in external magnetic fields}",
    eprint = "1206.4205",
    archivePrefix = "arXiv",
    primaryClass = "hep-lat",
    doi = "10.1103/PhysRevD.86.071502",
    journal = "Phys. Rev. D",
    volume = "86",
    pages = "071502",
    year = "2012"
}

@article{Bali:2014kia,
    author = {Bali, G. S. and Bruckmann, F. and Endr\"odi, G. and Katz, S. D. and Sch\"afer, A.},
    title = "{The QCD equation of state in background magnetic fields}",
    eprint = "1406.0269",
    archivePrefix = "arXiv",
    primaryClass = "hep-lat",
    doi = "10.1007/JHEP08(2014)177",
    journal = "Journal of High Energy Physics",
    volume = "08",
    pages = "177",
    year = "2014"
}

@article{Bali:2013owa,
    author = "Bali, G. S. and Bruckmann, F. and Endrodi, G. and Schafer, A.",
    title = "{Paramagnetic squeezing of QCD matter}",
    eprint = "1311.2559",
    archivePrefix = "arXiv",
    primaryClass = "hep-lat",
    doi = "10.1103/PhysRevLett.112.042301",
    journal = "Phys. Rev. Lett.",
    volume = "112",
    pages = "042301",
    year = "2014"
}

@article{Bali:2013esa,
    author = "Bali, G. S. and Bruckmann, F. and Endrodi, G. and Gruber, F. and Schaefer, A.",
    title = "{Magnetic field-induced gluonic (inverse) catalysis and pressure (an)isotropy in QCD}",
    eprint = "1303.1328",
    archivePrefix = "arXiv",
    primaryClass = "hep-lat",
    doi = "10.1007/JHEP04(2013)130",
    journal = "Journal of High Energy Physics",
    volume = "04",
    pages = "130",
    year = "2013"
}

@article{Bali:2012jv,
    author = "Bali, G. S. and Bruckmann, F. and Constantinou, M. and Costa, M. and Endrodi, G. and Katz, S. D. and Panagopoulos, H. and Schafer, A.",
    title = "{Magnetic susceptibility of QCD at zero and at finite temperature from the lattice}",
    eprint = "1209.6015",
    archivePrefix = "arXiv",
    primaryClass = "hep-lat",
    doi = "10.1103/PhysRevD.86.094512",
    journal = "Phys. Rev. D",
    volume = "86",
    pages = "094512",
    year = "2012"
}

@article{Bali:2020bcn,
    author = "Bali, Gunnar S. and Endr\H{o}di, Gergely and Piemonte, Stefano",
    title = "{Magnetic susceptibility of QCD matter and its decomposition from the lattice}",
    eprint = "2004.08778",
    archivePrefix = "arXiv",
    primaryClass = "hep-lat",
    doi = "10.1007/JHEP07(2020)183",
    journal = "Journal of High Energy Physics",
    volume = "07",
    pages = "183",
    year = "2020"
}

@article{Bruckmann:2013oba,
    author = "Bruckmann, Falk and Endrodi, Gergely and Kovacs, Tamas G.",
    title = "{Inverse magnetic catalysis and the Polyakov loop}",
    eprint = "1303.3972",
    archivePrefix = "arXiv",
    primaryClass = "hep-lat",
    doi = "10.1007/JHEP04(2013)112",
    journal = "Journal of High Energy Physics",
    volume = "04",
    pages = "112",
    year = "2013"
}

@article{Bornyakov:2013eya,
    author = {Bornyakov, V. G. and Buividovich, P. V. and Cundy, N. and Kochetkov, O. A. and Sch\"afer, A.},
    title = "{Deconfinement transition in two-flavor lattice QCD with dynamical overlap fermions in an external magnetic field}",
    eprint = "1312.5628",
    archivePrefix = "arXiv",
    primaryClass = "hep-lat",
    doi = "10.1103/PhysRevD.90.034501",
    journal = "Phys. Rev. D",
    volume = "90",
    pages = "034501",
    year = "2014"
}

@article{Endrodi:2015oba,
    author = "Endrodi, Gergely",
    title = "{Critical point in the QCD phase diagram for extremely strong background magnetic fields}",
    eprint = "1504.08280",
    archivePrefix = "arXiv",
    primaryClass = "hep-lat",
    doi = "10.1007/JHEP07(2015)173",
    journal = "Journal of High Energy Physics",
    volume = "07",
    pages = "173",
    year = "2015"
}

@article{DElia:2018xwo,
    author = "D'Elia, Massimo and Manigrasso, Floriano and Negro, Francesco and Sanfilippo, Francesco",
    title = "{QCD phase diagram in a magnetic background for different values of the pion mass}",
    eprint = "1808.07008",
    archivePrefix = "arXiv",
    primaryClass = "hep-lat",
    doi = "10.1103/PhysRevD.98.054509",
    journal = "Phys. Rev. D",
    volume = "98",
    pages = "054509",
    year = "2018"
}

@article{Ding:2020hxw,
    author = "Ding, H. -T. and Li, S. -T. and Tomiya, A. and Wang, X. -D. and Zhang, Y.",
    title = "{Chiral properties of (2+1)-flavor QCD in strong magnetic fields at zero temperature}",
    eprint = "2008.00493",
    archivePrefix = "arXiv",
    primaryClass = "hep-lat",
    doi = "10.1103/PhysRevD.104.014505",
    journal = "Phys. Rev. D",
    volume = "104",
    pages = "014505",
    year = "2021"
}

@article{Ding:2020inp,
    author = "Ding, Heng-Tong and Schmidt, Christian and Tomiya, Akio and Wang, Xiao-Dan",
    title = "{Chiral phase structure of three flavor QCD in a background magnetic field}",
    eprint = "2006.13422",
    archivePrefix = "arXiv",
    primaryClass = "hep-lat",
    doi = "10.1103/PhysRevD.102.054505",
    journal = "Phys. Rev. D",
    volume = "102",
    pages = "054505",
    year = "2020"
}

@article{Ding:2022tqn,
    author = "Ding, H. -T. and Li, S. -T. and Liu, J. -H. and Wang, X. -D.",
    title = "{Chiral condensates and screening masses of neutral pseudoscalar mesons in thermomagnetic QCD medium}",
    eprint = "2201.02349",
    archivePrefix = "arXiv",
    primaryClass = "hep-lat",
    doi = "10.1103/PhysRevD.105.034514",
    journal = "Phys. Rev. D",
    volume = "105",
    pages = "034514",
    year = "2022"
}

@article{DElia:2021yvk,
    author = "D'Elia, Massimo and Maio, Lorenzo and Sanfilippo, Francesco and Stanzione, Alfredo",
    title = "{Phase diagram of QCD in a magnetic background}",
    eprint = "2111.11237",
    archivePrefix = "arXiv",
    primaryClass = "hep-lat",
    doi = "10.1103/PhysRevD.105.034511",
    journal = "Phys. Rev. D",
    volume = "105",
    pages = "034511",
    year = "2022"
}

@article{Evans:2022hwr,
    author = "Evans, Geraint W. and Schmitt, Andreas",
    title = "{Chiral anomaly induces superconducting baryon crystal}",
    eprint = "2206.01227",
    archivePrefix = "arXiv",
    primaryClass = "hep-th",
    doi = "10.1007/JHEP09(2022)192",
    journal = "Journal of High Energy Physics",
    volume = "09",
    pages = "192",
    year = "2022"
}

@article{Brauner:2021sci,
    author = "Brauner, Tom\'a\v{s} and Kole\v{s}ov\'a, Helena and Yamamoto, Naoki",
    title = "{Chiral soliton lattice phase in warm QCD}",
    eprint = "2108.10044",
    archivePrefix = "arXiv",
    primaryClass = "hep-ph",
    doi = "10.1016/j.physletb.2021.136767",
    journal = "Phys. Lett. B",
    volume = "823",
    pages = "136767",
    year = "2021"
}

@article{Brauner:2016pko,
    author = "Brauner, Tomas and Yamamoto, Naoki",
    title = "{Chiral Soliton Lattice and Charged Pion Condensation in Strong Magnetic Fields}",
    eprint = "1609.05213",
    archivePrefix = "arXiv",
    primaryClass = "hep-ph",
    doi = "10.1007/JHEP04(2017)132",
    journal = "Journal of High Energy Physics",
    volume = "04",
    pages = "132",
    year = "2017"
}

@article{Adhikari:2022cks,
    author = "Adhikari, Prabal and Leeser, Elizabeth and Markowski, Jake",
    title = "{Phonon modes of magnetic vortex lattices in finite isospin chiral perturbation theory}",
    eprint = "2205.13369",
    archivePrefix = "arXiv",
    primaryClass = "hep-ph",
    doi = "10.1142/S0217732323500785",
    journal = "Mod. Phys. Lett. A",
    volume = "38",
    pages = "2350078",
    year = "2023"
}

@article{Adhikari:2018fwm,
    author = "Adhikari, Prabal",
    title = "{Magnetic Vortex Lattices in Finite Isospin Chiral Perturbation Theory}",
    eprint = "1810.03663",
    archivePrefix = "arXiv",
    primaryClass = "nucl-th",
    doi = "10.1016/j.physletb.2019.01.027",
    journal = "Phys. Lett. B",
    volume = "790",
    pages = "211--217",
    year = "2019"
}

@article{Adhikari:2015wva,
    author = "Adhikari, Prabal and Cohen, Thomas D. and Sakowitz, Julia",
    title = "{Finite Isospin Chiral Perturbation Theory in a Magnetic Field}",
    eprint = "1501.02737",
    archivePrefix = "arXiv",
    primaryClass = "nucl-th",
    doi = "10.1103/PhysRevC.91.045202",
    journal = "Phys. Rev. C",
    volume = "91",
    pages = "045202",
    year = "2015"
}

@article{Cohen:2013zja,
    author = "Cohen, Thomas D. and Yamamoto, Naoki",
    title = "{New critical point for QCD in a magnetic field}",
    eprint = "1310.2234",
    archivePrefix = "arXiv",
    primaryClass = "hep-ph",
    doi = "10.1103/PhysRevD.89.054029",
    journal = "Phys. Rev. D",
    volume = "89",
    pages = "054029",
    year = "2014"
}

@article{Shushpanov:1997sf,
    author = "Shushpanov, I. A. and Smilga, Andrei V.",
    title = "{Quark condensate in a magnetic field}",
    eprint = "hep-ph/9703201",
    archivePrefix = "arXiv",
    reportNumber = "ITEP-TH-6-97, TPI-MINN-97-04, NUC-MINN-97-2-T",
    doi = "10.1016/S0370-2693(97)00441-3",
    journal = "Phys. Lett. B",
    volume = "402",
    pages = "351--358",
    year = "1997"
}

@article{Agasian:1999sx,
    author = "Agasian, Nikita O. and Shushpanov, I. A.",
    title = "{The Quark and gluon condensates and low-energy QCD theorems in a magnetic field}",
    eprint = "hep-ph/9911254",
    archivePrefix = "arXiv",
    doi = "10.1016/S0370-2693(99)01414-8",
    journal = "Phys. Lett. B",
    volume = "472",
    pages = "143--149",
    year = "2000"
}

@article{Kharzeev:2012ph,
    author = "Kharzeev, Dmitri E. and Landsteiner, Karl and Schmitt, Andreas and Yee, Ho-Ung",
    title = "{Strongly interacting matter in magnetic fields}",
    eprint = "1211.6245",
    archivePrefix = "arXiv",
    primaryClass = "hep-ph",
    doi = "10.1007/978-3-642-37305-3",
    journal = "Lect. Notes Phys.",
    volume = "871",
    year = "2013"
}

@article{Miransky:2015ava,
    author = "Miransky, Vladimir A. and Shovkovy, Igor A.",
    title = "{Quantum field theory in a magnetic field: From quantum chromodynamics to graphene and Dirac semimetals}",
    eprint = "1503.00732",
    archivePrefix = "arXiv",
    primaryClass = "hep-ph",
    doi = "10.1016/j.physrep.2015.02.003",
    journal = "Phys. Rept.",
    volume = "576",
    pages = "1--209",
    year = "2015"
}

@article{Andersen:2014xxa,
    author = "Andersen, Jens O. and Naylor, William R. and Tranberg, Anders",
    title = "{Phase diagram of QCD in a magnetic field: A review}",
    eprint = "1411.7176",
    archivePrefix = "arXiv",
    primaryClass = "hep-ph",
    doi = "10.1103/RevModPhys.88.025001",
    journal = "Rev. Mod. Phys.",
    volume = "88",
    pages = "025001",
    year = "2016"
}

@article{Coppola:2019idh,
    author = "Coppola, Maximo and Gomez Dumm, Daniel and Noguera, Santiago and Scoccola, Norberto N.",
    title = "{Magnetic field driven enhancement of the weak decay width of charged pions}",
    eprint = "1908.10765",
    archivePrefix = "arXiv",
    primaryClass = "hep-ph",
    doi = "10.1007/JHEP09(2020)058",
    journal = "Journal of High Energy Physics",
    volume = "09",
    pages = "058",
    year = "2020"
}

@article{Husek:2024gnn,
    author = "Husek, Tom\'a\v{s}",
    title = "{The neutral-pion decay into electron-positron pair: A review and update}",
    eprint = "2405.09650",
    archivePrefix = "arXiv",
    primaryClass = "hep-ph",
    journal="",
    year = "2024"
}

@article{Miransky:2002rp,
    author = "Miransky, V. A. and Shovkovy, I. A.",
    title = "{Magnetic catalysis and anisotropic confinement in QCD}",
    eprint = "hep-ph/0205348",
    archivePrefix = "arXiv",
    doi = "10.1103/PhysRevD.66.045006",
    journal = "Phys. Rev. D",
    volume = "66",
    pages = "045006",
    year = "2002"
}

@article{Coppola:2019uyr,
    author = "Coppola, M. and Gomez Dumm, D. and Noguera, S. and Scoccola, N. N.",
    title = "{Neutral and charged pion properties under strong magnetic fields in the  NJL model}",
    eprint = "1907.05840",
    archivePrefix = "arXiv",
    primaryClass = "hep-ph",
    doi = "10.1103/PhysRevD.100.054014",
    journal = "Phys. Rev. D",
    volume = "100",
    pages = "054014",
    year = "2019"
}

@article{Coppola:2019wvh,
    author = "Coppola, M. and Gomez Dumm, D. and Noguera, S. and Scoccola, N. N.",
    title = "{Weak decays of magnetized charged pions in the symmetric gauge}",
    eprint = "1910.10814",
    archivePrefix = "arXiv",
    primaryClass = "hep-ph",
    doi = "10.1103/PhysRevD.101.034003",
    journal = "Phys. Rev. D",
    volume = "101",
    pages = "034003",
    year = "2020"
}

@article{Coppola:2018ygv,
    author = "Coppola, M. and Gomez Dumm, D. and Noguera, S. and Scoccola, N. N.",
    title = "{Pion-to-vacuum vector and axial vector amplitudes and weak decays of pions in a magnetic field}",
    eprint = "1810.08110",
    archivePrefix = "arXiv",
    primaryClass = "hep-ph",
    doi = "10.1103/PhysRevD.99.054031",
    journal = "Phys. Rev. D",
    volume = "99",
    pages = "054031",
    year = "2019"
}

@article{Adler:1969gk,
    author = "Adler, Stephen L.",
    title = "{Axial vector vertex in spinor electrodynamics}",
    doi = "10.1103/PhysRev.177.2426",
    journal = "Phys. Rev.",
    volume = "177",
    pages = "2426--2438",
    year = "1969"
}

@article{Bell:1969ts,
    author = "Bell, J. S. and Jackiw, R.",
    title = "{A PCAC puzzle: $\pi^0 \to \gamma \gamma$ in the $\sigma$ model}",
    doi = "10.1007/BF02823296",
    journal = "Nuovo Cim. A",
    volume = "60",
    pages = "47--61",
    year = "1969"
}

@article{Wess:1971yu,
    author = "Wess, J. and Zumino, B.",
    title = "{Consequences of anomalous Ward identities}",
    doi = "10.1016/0370-2693(71)90582-X",
    journal = "Phys. Lett. B",
    volume = "37",
    pages = "95--97",
    year = "1971"
}

@article{Witten:1983tw,
    author = "Witten, Edward",
    title = "{Global Aspects of Current Algebra}",
    reportNumber = "PRINT-83-0262 (PRINCETON)",
    doi = "10.1016/0550-3213(83)90063-9",
    journal = "Nucl. Phys. B",
    volume = "223",
    pages = "422--432",
    year = "1983"
}

@article{Tiburzi:2008ma,
    author = "Tiburzi, Brian C.",
    title = "{Hadrons in Strong Electric and Magnetic Fields}",
    eprint = "0808.3965",
    archivePrefix = "arXiv",
    primaryClass = "hep-ph",
    reportNumber = "UMD-40762-426",
    doi = "10.1016/j.nuclphysa.2008.10.010",
    journal = "Nucl. Phys. A",
    volume = "814",
    pages = "74--108",
    year = "2008"
}

@article{Bijnens:2014lea,
    author = "Bijnens, Johan and Ecker, Gerhard",
    title = "{Mesonic Low-Energy Constants}",
    eprint = "1405.6488",
    archivePrefix = "arXiv",
    primaryClass = "hep-ph",
    reportNumber = "LU-TP-14-16, UWTHPH-2014-10",
    doi = "10.1146/annurev-nucl-102313-025528",
    journal = "Ann. Rev. Nucl. Part. Sci.",
    volume = "64",
    pages = "149--174",
    year = "2014"
}

@article{Kaiser:2000ck,
    author = "Kaiser, Roland",
    title = "{Anomalies and WZW Term of Two-Flavor QCD}",
    eprint = "hep-ph/0011377",
    archivePrefix = "arXiv",
    reportNumber = "BUTP-00-23",
    doi = "10.1103/PhysRevD.63.076010",
    journal = "Phys. Rev. D",
    volume = "63",
    pages = "076010",
    year = "2001"
}

@article{Gasser:1984gg,
    author = "Gasser, J. and Leutwyler, H.",
    title = "{Chiral Perturbation Theory: Expansions in the Mass of the Strange Quark}",
    reportNumber = "CERN-TH-3798",
    doi = "10.1016/0550-3213(85)90492-4",
    journal = "Nucl. Phys. B",
    volume = "250",
    pages = "465--516",
    year = "1985"
}

@article{Gasser:1983yg,
    author = "Gasser, J. and Leutwyler, H.",
    title = "{Chiral Perturbation Theory to One Loop}",
    reportNumber = "CERN-TH-3689",
    doi = "10.1016/0003-4916(84)90242-2",
    journal = "Annals Phys.",
    volume = "158",
    pages = "142",
    year = "1984"
}

@article{Adhikari:2023fdl,
    author = "Adhikari, Prabal and Tiburzi, Brian C.",
    title = "{QCD thermodynamics and neutral pion in a uniform magnetic field: Finite volume effects}",
    eprint = "2302.09179",
    archivePrefix = "arXiv",
    primaryClass = "hep-lat",
    doi = "10.1103/PhysRevD.107.094504",
    journal = "Phys. Rev. D",
    volume = "107",
    pages = "094504",
    year = "2023"
}

@article{Werbos:2007ym,
    author = "Werbos, Elizabeth S.",
    title = "{The Chiral condensate in a constant electromagnetic field at $O(p^6)$}",
    eprint = "0711.2635",
    archivePrefix = "arXiv",
    primaryClass = "hep-ph",
    doi = "10.1103/PhysRevC.77.065202",
    journal = "Phys. Rev. C",
    volume = "77",
    pages = "065202",
    year = "2008"
}

@article{Adhikari:2021lbl,
    author = "Adhikari, Prabal",
    title = "{Topological susceptibility in a uniform magnetic field}",
    eprint = "2103.05048",
    archivePrefix = "arXiv",
    primaryClass = "hep-ph",
    doi = "10.1016/j.physletb.2021.136826",
    journal = "Phys. Lett. B",
    volume = "825",
    pages = "136826",
    year = "2022"
}

@article{Adhikari:2021jff,
    author = "Adhikari, Prabal",
    title = "{QCD \ensuremath{\theta}-vacuum in a uniform magnetic field}",
    eprint = "2111.06196",
    archivePrefix = "arXiv",
    primaryClass = "hep-ph",
    doi = "10.1016/j.nuclphysb.2021.115627",
    journal = "Nucl. Phys. B",
    volume = "974",
    pages = "115627",
    year = "2022"
}

@article{Tiburzi:2014zva,
    author = "Tiburzi, Brian C.",
    title = "{Neutron in a Strong Magnetic Field: Finite Volume Effects}",
    eprint = "1403.0878",
    archivePrefix = "arXiv",
    primaryClass = "hep-lat",
    reportNumber = "RBRC-1061",
    doi = "10.1103/PhysRevD.89.074019",
    journal = "Phys. Rev. D",
    volume = "89",
    pages = "074019",
    year = "2014"
}

@article{Deshmukh:2017ciw,
    author = "Deshmukh, Amol and Tiburzi, Brian C.",
    title = "{Octet Baryons in Large Magnetic Fields}",
    eprint = "1709.04997",
    archivePrefix = "arXiv",
    primaryClass = "hep-ph",
    doi = "10.1103/PhysRevD.97.014006",
    journal = "Phys. Rev. D",
    volume = "97",
    pages = "014006",
    year = "2018"
}

@article{Andersen:2012zc,
    author = "Andersen, Jens O.",
    title = "{Chiral perturbation theory in a magnetic background - finite-temperature effects}",
    eprint = "1205.6978",
    archivePrefix = "arXiv",
    primaryClass = "hep-ph",
    doi = "10.1007/JHEP10(2012)005",
    journal = "Journal of High Energy Physics",
    volume = "10",
    pages = "005",
    year = "2012"
}

@article{Andersen:2012dz,
    author = "Andersen, Jens O.",
    title = "{Thermal pions in a magnetic background}",
    eprint = "1202.2051",
    archivePrefix = "arXiv",
    primaryClass = "hep-ph",
    doi = "10.1103/PhysRevD.86.025020",
    journal = "Phys. Rev. D",
    volume = "86",
    pages = "025020",
    year = "2012"
}

@article{Urech:1994hd,
    author = "Urech, Res",
    title = "{Virtual Photons in Chiral Perturbation Theory}",
    eprint = "hep-ph/9405341",
    archivePrefix = "arXiv",
    reportNumber = "BUTP-94-9",
    doi = "10.1016/0550-3213(95)90707-N",
    journal = "Nucl. Phys. B",
    volume = "433",
    pages = "234--254",
    year = "1995"
}

@article{Bijnens:1999sh,
    author = "Bijnens, Johan and Colangelo, Gilberto and Ecker, Gerhard",
    title = "{The Mesonic chiral Lagrangian of order $p^6$}",
    eprint = "hep-ph/9902437",
    archivePrefix = "arXiv",
    reportNumber = "LU-TP-99-02, UWTHPH-1999-02, ZU-TH-9-99",
    doi = "10.1088/1126-6708/1999/02/020",
    journal = "Journal of High Energy Physics",
    volume = "02",
    pages = "020",
    year = "1999"
}

@article{Cirigliano:2006hb,
    author = "Cirigliano, V. and Ecker, G. and Eidemuller, M. and Kaiser, Roland and Pich, A. and Portoles, J.",
    title = "{Towards a consistent estimate of the chiral low-energy constants}",
    eprint = "hep-ph/0603205",
    archivePrefix = "arXiv",
    reportNumber = "CALTECH-MAP-319, CPT-P05-2006, FTUV-06-0319, IFIC-06-02, UWTHPH-2005-19",
    doi = "10.1016/j.nuclphysb.2006.07.010",
    journal = "Nucl. Phys. B",
    volume = "753",
    pages = "139--177",
    year = "2006"
}

@article{Cohen:2007bt,
    author = "Cohen, Thomas D. and McGady, David A. and Werbos, Elizabeth S.",
    title = "{The Chiral condensate in a constant electromagnetic field}",
    eprint = "0706.3208",
    archivePrefix = "arXiv",
    primaryClass = "hep-ph",
    doi = "10.1103/PhysRevC.76.055201",
    journal = "Phys. Rev. C",
    volume = "76",
    pages = "055201",
    year = "2007"
}

@article{Bali:2018sey,
    author = {Bali, G. S. and Brandt, B. B. and Endr\H{o}di, G. and Gl\"a\ss{}le, B.},
    title = "{Weak decay of magnetized pions}",
    eprint = "1805.10971",
    archivePrefix = "arXiv",
    primaryClass = "hep-lat",
    doi = "10.1103/PhysRevLett.121.072001",
    journal = "Phys. Rev. Lett.",
    volume = "121",
    pages = "072001",
    year = "2018"
}

@article{Bali:2017ian,
    author = {Bali, Gunnar S. and Brandt, Bastian B. and Endr\H{o}di, Gergely and Gl\"a\ss{}le, Benjamin},
    title = "{Meson masses in electromagnetic fields with Wilson fermions}",
    eprint = "1707.05600",
    archivePrefix = "arXiv",
    primaryClass = "hep-lat",
    doi = "10.1103/PhysRevD.97.034505",
    journal = "Phys. Rev. D",
    volume = "97",
    pages = "034505",
    year = "2018"
}

@article{Harada:2003jx,
    author = "Harada, Masayasu and Yamawaki, Koichi",
    title = "{Hidden local symmetry at loop: A New perspective of composite gauge boson and chiral phase transition}",
    eprint = "hep-ph/0302103",
    archivePrefix = "arXiv",
    reportNumber = "DPNU-03-02",
    doi = "10.1016/S0370-1573(03)00139-X",
    journal = "Phys. Rept.",
    volume = "381",
    pages = "1--233",
    year = "2003"
}

@article{THaugset:1993,
doi = {10.1088/0031-8949/47/6/004},
year = {1993},
volume = {47},
pages = {715},
author = {T Haugset and  J Aa Ruud and  F Ravndal},
title = "{Gauge invariance of Landau levels}",
journal = {Physica Scripta}
}

\end{document}